\documentclass[onecolumn,nofootinbib,preprintnumbers, sort&compress, floatfix,a4paper,superscriptaddress,9pt]{revtex4-1}
\usepackage[framemethod=TikZ]{mdframed}
\usepackage{xcolor}
\usepackage{amsthm}
\usepackage{pgf}
\usepackage{pgfpages}
\usepackage{amsmath}
\usepackage{amsfonts}
\usepackage{amssymb}
\usepackage{graphicx}
\usepackage{bm}
\usepackage{hyperref}
\usepackage{lipsum}
\usepackage{array}
\usepackage{booktabs}
\usepackage{tikz}
\usepackage{titlesec}
\usepackage{hyperref}
\hypersetup{bookmarksnumbered}
\def\be{\begin{equation}}
\def\ee{\end{equation}}
\def\bi{\bibitem}
\usepackage{amsmath}
\usepackage{empheq}
\usepackage[most]{tcolorbox}
\newtcbox{\mymath}[1][]{%
    nobeforeafter, math upper, tcbox raise base,
    enhanced, colframe=blue!30!black,
    colback=blue!30, boxrule=1pt,
    #1}
\usepackage{mdframed}
\usepackage{lipsum}
\usepackage{PGF}
\usepackage{pgfkeys}
\usepackage{pgffor}
\usepackage{pgfcalendar}
\usepackage{pgfpages}
\usepackage{tikz}
\usetikzlibrary{calc}
\usepackage{array}
\usepackage{booktabs}
\usepackage{multirow}
\usepackage{latexsym}
\usepackage[english]{babel}
\allowdisplaybreaks
\hypersetup{colorlinks=true,linkcolor=blue!100,citecolor=red!100}
\newcounter{theo}[section] 
\renewcommand{\thetheo}{\arabic{section}.\arabic{theo}}

\newcounter{prf}[section]
\renewcommand{\theprf}{\arabic{section}.\arabic{prf}}

\newcounter{lem}[section]
\renewcommand{\thelem}{\arabic{section}.\arabic{lem}}

\titleformat{\section}
{\color{black}\normalfont\Large\bfseries}
{\color{black}\thesection}{1em}{}
\renewcommand\thesection{\arabic{section}}

\newcommand*{\Rrmybox}[2]{\colorbox[rgb]{1.00,0.47,1.00}{\parbox{.99\linewidth}{#2}}}
\newcommand*{\hrmybox}[2]{\colorbox[rgb]{0.70,0.70,1.00}{\parbox{.99\linewidth}{#2}}}
\newcommand*{\rmybox}[2]{\colorbox[rgb]{0.57,0.78,1.00}{\parbox{.99\linewidth}{#2}}}
\newcommand*{\qrmybox}[2]{\colorbox[rgb]{0.46,0.89,0.49}{\parbox{.99\linewidth}{#2}}}
\newcommand*{\pqrmybox}[2]{\colorbox[rgb]{0.00,0.98,0.25}{\parbox{.99\linewidth}{#2}}}
\newcommand*{\goldmybox}[2]{\colorbox[rgb]{0.76,0.98,0.25}{\parbox{.99\linewidth}{#2}}}
\newcommand*{\rgoldmybox}[2]{\colorbox[rgb]{0.00,0.82,0.82}{\parbox{.99\linewidth}{#2}}}
\allowdisplaybreaks
\begin{document}
\title{\textcolor[rgb]{0.00,0.40,0.80}{\textsf{\huge\boldmath Reconstruction of \(F(T)\) gravity in homogeneous backgrounds.}}}

\author{‪Behzad Tajahmad}
\email{behzadtajahmad@yahoo.com}
\affiliation{Faculty of Physics, University of Tabriz, Tabriz, Iran}
\affiliation{Research Institute for Astronomy and Astrophysics of Maragha (RIAAM)-Maragha, Iran, P.O. Box: 55134-441\\}
\begin{center}
\begin{abstract}
\begin{tcolorbox}[breakable,colback=white,
colframe=cyan,width=\dimexpr\textwidth+0mm\relax,enlarge left by=-17mm,enlarge right by=-6mm ]\normalsize{\textbf{\textsf{Abstract:}} A technique for the reconstruction of the potential for a scalar field in cosmological models based on induced gravity has recently been developed by Alexander Y. Kamenshchik, Alessandro Tronconi, and Giovanni Venturi \cite{ref1}. In this paper, this extended reconstruction method is utilized to investigate the nature of the scalar potentials of the most extended action of \(F(T)\)-gravity context. First, a general formalism is formulated and then it is utilized for considering some well-known special cases including `Barotropic Fluid', `Cosmological Constant', and `Modified Chaplygin Gas'.\\
The analysis of the results is carried out by the use of the $\mathfrak{B}\text{-function}$ method which has recently been suggested by the author \cite{hashem}.\\
As we know, assuming a proportional relation between the scale factor of the $x\text{-direction}$ and the scale factors of $y$ and $z$ directions (i.e. $A=B^{m}$) in dealing with LRS Bianchi-I background is prevalent. Pursuant to observational data, the correct physical range of $m$ is extracted. It is demonstrated that, unlike several papers, $m$ is very close to $1$.\\
Some interesting discussions about the modified and generalized Chaplygin gases are performed. The ranges of the amounts of free parameters of the various types of Chaplygin gases are corrected according to observational data. It is demonstrated that the generalized Chaplygin gas model namely $P=-\sigma_{2}/\rho^{\nu}$ may be developed as $P=-\sigma_{2}/\rho^{\nu(t)}$ (i.e. $\nu=const. \longrightarrow \nu=\nu(t)$).\\
Ultimately, by combining ref. \cite{ref1} and the current paper, a general prescription is recommended for the reconstruction of the potentials of alternative theories of gravity (especially \(F(R)\) and \(F(T)\)), and their relevant analysis.} \end{tcolorbox}
\end{abstract}

\end{center}
\maketitle
\break
\hrule \hrule \hrule
\textbf{\textcolor[rgb]{0.00,0.00,0.00}{\tableofcontents}}
\text{ }
\hrule \hrule \hrule
\noindent\hrmybox{}{\section{Introduction\label{sec:intro}}}\vspace{5mm}

Explanation of the essence and mechanism of the acceleration of our universe, proven by several strong astronomical and cosmological observations including supernova type Ia \cite{sup1,sup2}, CMB studies \cite{CMB}, weak lensing \cite{lens}, baryon acoustic oscillations \cite{baryon}, and large-scale structure \cite{LSC}, is one of the great and major challenges for physicists.

The accelerating expansion of the universe is driven by so-called dark energy which is a mysterious energy with negative pressure \cite{DE1}. Two major problems like `fine tuning' and `cosmic coincidence' are related to dark energy. The most probable solution to dark energy was thought that is the Einstein's cosmological constant \cite{Sol-DE}, but it failed because it cannot resolve
the two problems mentioned above. Hence, other feasible theoretical models by considering the dynamic nature of dark energy like the phantom field \cite{phantom1, phantom2, phantom3, phantom4, phantom5, phantom6}, quintessence \cite{quin1, quin2, quin3}, quintom \cite{quintom1, quintom2, quintom3, quintom4}, tachyon field \cite{tachyon}, and the interacting dark energy models like holographic models \cite{holo1,holo2}, Chaplygin gas \cite{Chap}, braneworld models \cite{brane} and etcetera, have been suggested to interpret our accelerating universe. Another possibility is to modify Einstein's general relativity (Modified Gravity) \cite{MG1, MG2}. Indeed, in this approach, the action of the theory is made dependent on a function of the curvature scalar. As we anticipate, in a certain limit of the parameters, the theory reduces to general relativity. Theories like $F(R)\text{-gravity}$, $F(T)\text{-gravity}$, $F(T)\text{-gravity}$ with an unusual term \cite{beh-intervention}, and scalar-tensor theories are the fruits of many attempts to this modification by physicists \cite{lio}. This new set of gravity theories passes several solar system and astrophysical tests successfully \cite{solar1,solar2,solar3,vasilisok}.

Teleparallel gravity is a gravity theory which uses the curvature-free Weitzenb\"{o}ck connection to define the covariant derivative, instead of the conventional torsion-less Levi-Civita connection of general relativity, and attempts to describe the effects of gravitation in terms of torsion in lieu of curvature. In order to the sake of unifying gravity and electromagnetism, teleparallel gravity was initially introduced by Einstein. Though it is equivalent to general relativity in its simplest form, but it has a different physical interpretation. The field equations in this theory are second-order differential equations, while for the generalized $F(R)$ theory they are of the fourth order, hence it, admittedly, is simpler to analyze. One of the modifications of the matter part of the Einstein-Hilbert action is $F(T)$ gravity as an extension of teleparallel gravity. $F(T)$ gravity has recently received attention in the literature \cite{ft1,ft2,ft3,ft4,ft5,ft6,ft7,ft8,ft9,ft10,ft11,ft12,ft13,ft14,ft15,ft16,
ft17,ft18,ft19,ft20}, mostly in the context of explaining the observed acceleration of the universe.

In unified theories of interactions and also in inflationary scenarios in cosmology, scalar fields have a substantial role. Indeed, a rich variety of dark energy and inflationary models may be accommodated phenomenologically by scalar fields in which the inflations produce the initial acceleration. Hence, the technique of the reconstruction of the potentials of scalar fields has been taken into account. This technique enables one to find the form of the scalar field potential as well as the scalar field for a particular value of the Hubble parameter in terms of scale factor or cosmic time, or particularly the redshift. For instance, in the current paper, the Hubble parameter arisen from `Barotropic fluid', `Cosmological constant', and `Modified Chaplygin gas' are investigated. Furthermore, the reconstruction of scalar-field dark energy models from observations has attracted the attention of researchers for a long time \cite{367, 368, 369, 370, 371}. In fact, one can reconstruct the potential and the equation of state of the field by parameterizing the Hubble parameter $H$ in terms of the redshift $z$ \cite{372}. The Hubble rate $H(z)$ is determined by the luminosity distance $d_L(z)$ by using the relation
\begin{align*}
H(z)=\left[\frac{\mathrm{d}}{\mathrm{d}z}\left(\frac{d_{L}(z)}{1+z} \right) \right]^{-1}.
\end{align*}
If the luminosity distance observationally is measured, the expansion rate of the universe is determined. This method was generalized to scalar-tensor theories \cite{354, 355, 373.1, 373.2}, $F(R)$ gravity \cite{374,venus1r,venus2r,venus3r}, and also a dark-energy fluid with viscosity terms \cite{375}. Recently, a bottom-up \(F(R)\) gravity reconstruction technique has also been introduced by S.D. Odintsov and V.K. Oikonomou \cite{1368}.\\
In this paper, we apply a technique for the reconstruction of the potential for a scalar field in $F(T)$-gravity context, which has been developed in cosmological models based on induced gravity by Alexander Y. Kamenshchik, Alessandro Tronconi, and Giovanni Venturi \cite{ref1}. However, prior to ref. \cite{ref1}, the reconstruction of scalar theory (actually potentials) for different evolutions was given in \cite{venus1p,venus2p}, but note that there are some differences among these approaches with \cite{ref1}.

Recently, a new approach to the analysis of the reconstruction methods, phase space, and exact solutions of the alternative theories of gravity has been suggested \cite{hashem}. The results of this paper are analyzed through this new method based on ``Class-1'' (Decreasing scalar fields with time). It is strongly recommended that the reader first study the ref. \cite{hashem} carefully, otherwise, the current paper will be obscure.\\

\noindent\hrmybox{}{\section{The Geometry of Background}}\vspace{5mm}

In this paper, the homogeneous backgrounds (i.e. FRW and Bianchi type) are desired for investigating the extended action of $f(T)$ gravity. Hence, we start with the LRS BI (Locally  Rotationally Symmetric Bianchi type I) universe model which is given by
\begin{align}\label{line element}
\mathrm{d}s^2=\mathrm{d}t^2-A^2(t)\mathrm{d}x^2-B^2(t) \left[\mathrm{d}y^2+\mathrm{d}z^2\right],
\end{align}
where the expansion radii $A$ and $B$ are functions of time $t$.
Therefore, the torsion scalar for this background would be
\begin{align}\label{torsion}
T= -2 \left(2\frac{\dot{A}}{A}\frac{\dot{B}}{B} +\frac{\dot{B}^2}{B^2} \right)
=-2 \left(2H_{1}H_{2}+H^2_{2}\right),
\end{align}
where the dot indicates a differentiation with respect to time and $H_{1}$, and $H_{2}$ are the directional Hubble parameters ($H_{1}$ along $x$ direction while $H_{2}$ along $y$ and $z$ directions). Before proceeding, let us restrict ourselves to a well-known physical assumption: $A=B^{m}$ with $m\neq 0$. Note that $m=0$ is nonphysical because it means that one of the scale factors is constant (i.e. $A=1$), and $m=1$ is FRW space-time. This relation arises from the condition that in a spatially homogeneous model the ratio of shear scalar $\sigma$,
\begin{align}\label{zub-sig}
\sigma^2=\frac{1}{2}\sigma_{ab}\sigma^{ab}=\frac{1}{3}
\left(\frac{\dot{A}}{A}-\frac{\dot{B}}{B} \right)^2,
\end{align}
to expansion scalar $\Theta$,
\begin{align}\label{zub-Thet}
\Theta=u^{a}_{;a}=\frac{\dot{A}}{A}+2\frac{\dot{B}}{B},
\end{align}
is constant (i.e. $\sigma / \Theta=constant$). Using the condition $A=B^{m}$, the torsion scalar (\ref{torsion}), the shear scalar (\ref{zub-sig}), and the expansion scalar (\ref{zub-Thet}) take the forms
\begin{align}
&T=-2(2m+1) \frac{\dot{B}^2}{B^2}=-2(2m+1)H^2_{2}, \label{torsion1}\\
&\sigma^2=\frac{(m-1)^2}{3}H^{2}_{2}, \label{zub5}
\end{align}
and
\begin{align}
\Theta=(m+2)H_{2}, \label{zub4}
\end{align}
respectively. In the current paper, we take care of calculations to be general and correct for both FRW ($m=1$) and LRS BI ($m \neq 1$) geometrical backgrounds, but in the case that the formula cannot be held for both cases, then we consider it separately.\\
Pursuing the background geometry (\ref{line element}) under the condition $A=B^m$, the average scale factor, the volume, and the average Hubble parameter are defined as
\begin{align}\label{zub1}
a \equiv a_{\mathrm{ave.}} &=\left(AB^2 \right)^{\frac{1}{3}}=B^{\frac{(m+2)}{3}}, \quad \text{\textit{Vol.}}=a^3_{\mathrm{ave.}}=B^{(m+2)}, \nonumber \\
H \equiv H_{\mathrm{ave.}} &=\frac{(m+2)}{3}H_{2}.
\end{align}
The anisotropy parameter of the expansion is characterized by the mean ($H$) and directional Hubble parameters ($H_{i}$), and it is defined as
\begin{align}\label{zub2}
\Delta = \frac{1}{3}\sum_{i=1}^{3}\left(\frac{H_{i}-H}{H} \right)^2=2\frac{(m-1)^2}{(m+2)^2}.
\end{align}
Therefore, the anisotropy parameter of the expansion and the shear scalar become zero at $m=1$ (Flat FRW universe).

Now, let us perform some simple calculations to define exactly the physically admissible range of $m$ according to observational data. The condition $(\sigma / H)\leq N_{0}$, where $N_{0}$ is about $10^{-10}$ (See ref. \cite{PRL}; and note that total shear obeys $\sigma^2=\sigma^2_{\text{Scalar mode}}+\sigma^2_{\text{Vector mode}}+\sigma^2_{\text{Tensor mode}}$), leads to
\begin{align}\label{zub6}
\frac{\sigma}{\hat{H}}=\frac{\sqrt{3}\left|m-1 \right|}{(m+2)}\leq N_{0}
\end{align}
So, one has
\begin{equation}\begin{split}\label{zub6}
\begin{aligned}
\left\{
\begin{array}{ll}
\text{if }\; m \geq 1: \quad &{}\; 1 \leq m \leq \left(\frac{\sqrt{3}+2N_{0}}{\sqrt{3}-N_{0}} \right)\equiv N^{+}_{0}\\  \\
\text{if }\; 0<m< 1: &{}\; N^{-}_{0} \equiv \left(\frac{\sqrt{3}-2N_{0}}{\sqrt{3}+N_{0}} \right) \leq m <1
\end{array} \right.
\end{aligned}\end{split}\end{equation}
Note that we have focused on the positive values of $m$ since negative amounts of $m$ are nonphysical; in ref. \cite{hashem}, this fact has been proved. Combining both conditions yields
\begin{align}\label{m-range}
\boxed {N^{-}_{0}\leq m \leq N^{+}_{0}}
\end{align}
As we observe, $m$ is very close to $1$, and as a result, $\Delta$ would be very close to zero.\\

By assuming that the cosmic matter is represented by the energy-momentum tensor of a perfect fluid
\begin{align}\label{T}
T_{ij}=\left(\rho+P\right)v_{i}v_{j}+Pg_{ij},
\end{align}
where $\rho$ is the energy density of the cosmic matter, $P$ is its pressure, and $v_{i}$ is the four-velocity vector such that $v_{i}v^{i}=1$,
the Einstein field equations would be
\begin{align}
2\frac{\ddot{B}}{B}+\frac{\dot{B}^2}{B^2}=-8\pi GP+\Lambda, \label{ms21}\\
\frac{\ddot{A}}{A}+\frac{\ddot{B}}{B}+\frac{\dot{A}}{A}\frac{\dot{B}}{B}=-8\pi GP+\Lambda, \label{ms22}\\
\frac{\ddot{A}}{A}+\frac{\ddot{B}}{B}+\frac{\dot{A}}{A}\frac{\dot{B}}{B}=-8\pi GP+\Lambda, \label{ms23}\\
2\frac{\dot{A}}{A}\frac{\dot{B}}{B}+\frac{\dot{B}^2}{B^2}=8\pi G\rho+\Lambda. \label{ms24}
\end{align}
In this paper, we set $8\pi G\equiv 1$ and $\Lambda=0$. The usual energy conservation equation $T^j_{i;j}=0$ yields
\begin{align}\label{ms3}
\dot{\rho}+\left(\rho+P \right)\left(\frac{\dot{A}}{A}
+2\frac{\dot{B}}{B}\right)=0.
\end{align}
Under the physical assumption $A=B^m$, eq. (\ref{ms3}) reads
\begin{align}\label{chegali}
\dot{\rho}+(m+2)\left(\rho+P\right)H_{2}=0.
\end{align}
This equation is known as the ``conservation'' equation for the LRS BI universe. Summing eqs. (\ref{ms21})-(\ref{ms23}) lead to
\begin{align}\label{bikhod}
(2m^2+2m+5)H^2_{2}+(2m+4)\dot{H}_{2}=-3P.
\end{align}
This equation may be written as
\begin{align}\label{hard1}
\left(\frac{m^2+m+4}{2}\right)H^2_{2}+\left(\frac{m+3}{2}\right)
\dot{H}_{2}= -P.
\end{align}
And, eq. (\ref{ms24}) simply reads
\begin{align}\label{hard2}
(2m+1)H^2_{2}=\rho.
\end{align}
Eqs. (\ref{hard1}) and (\ref{hard2}) are the Einstein field equations for LRS BI universe model (Under the condition $A=B^m$).\\

\noindent\hrmybox{}{\section{The Model\label{II}}}\vspace{5mm}

A comprehensive treatment of the problem of current interest, namely generalization of the $F(T)$ term in general relativity, has recently been done \cite{beh-intervention} through offering the following action
\begin{align}\label{action}
S = \int d^4x \; e \bigg[ f(\varphi)T &- U\left(\varphi ,\varphi_{,\mu}\varphi^{,\mu} \right) g(T) + \frac{\omega(\varphi)}{2}\varphi_{,\mu} \varphi^{,\mu} - V(\varphi) \bigg],
\end{align}
where $e=\det(e_{\nu}^{i})=\sqrt{-g}$ with $e_{\nu}^{i}$ being a vierbein (tetrad) basis, $f(\varphi)$ is the generic function describing the coupling between the scalar field and the scalar torsion $T$, $\varphi_{,\mu}$ represents the covariant derivative of $\varphi$, $U\left(\varphi ,\varphi_{,\mu}\varphi^{,\mu} \right)$ is the unknown coupling function which has been hypothesized, in general, to depend upon the scalar field and its gradients. This function has been coupled with an unknown function of torsion $g(T)$. Here, $\omega(\varphi)$ and $V(\varphi)$ are the coupling function and the scalar potential, respectively. It is worth mentioning that the scalars here are caused by conformal symmetry \cite{odin1}. Indeed, the generic action of teleparallel gravity has been extended by adding the new (unusual) term (i.e. $U\left(\varphi ,\varphi_{,\mu}\varphi^{,\mu} \right) g(T)$). Under certain ans\"{a}tz, this model led to quite acceptable agreements with observational details and Noether's theorem used to compute the symmetries and consequently the conservation laws (See Ref. \cite{beh-intervention}).

It is worth trying to employ the reconstruction method for investigating the nature of the scalar potentials. However, our complicated action implies making some changes in the reconstruction procedure proposed by A.Y. Kamenshchik et al. (See Ref. \cite{ref1}). According to another paper, ref. \cite{Waheed}, it may be argued that there is no general basic equation for all the cases of interest in \(F(R)\)-gravity, nonetheless, we intend to make this technique applicable to our model and ultimately, obtain a general basic formalism such that it works at least for some well-known cases.\\
First of all, knowing the form of $g(T)$ is necessary for earning a general formalism. In continuation of the paper, it is shown that by the form of $g(T)$ at hand, one is able to formulate a general procedure. There is a physical point behind it: ``The key point of the reconstruction technique is the assumption that Hubble parameter is dependent on the scalar field $\varphi$. Therefore, the scalar field has the chief role and within this method, it is linked to the scale factor instead of time. On the other hand, altogether, each unknown function in the action (\ref{action}), excluding $g(T)$, feeds on the scalar field and its derivatives, so $g(T)$ must be replaced by its form in which it depends explicitly upon the Hubble parameter.''. But, `Which form do we choose for $g(T)$?'. Before answering this question, it must be stated that in the current paper, the cases $f(\varphi)=f_{0}= constant$, and $f(\varphi)=\xi \varphi^2$ are noteworthy. Both special cases are perused by taking three different matter contents, namely `Barotropic fluid',  `Cosmological constant', and `Modified Chaplygin gas', into account. However, we first formulate a general formalism (by replacing a suitable function for $g(T)$) and then study these cases, but when we investigate these special forms, the unknown forms of the other unknown functions, excluding the potential, are problematic. Hence, it is indispensable to utilize a standard approach such as Noether symmetry approach --- it is a suggestion not more --- to determine the forms of $U$, $g$, and $\omega$ provided that $f(\varphi)$ be $f_{0}$, and $\xi \varphi^2$, and then presenting the scrutiny of special cases would be feasible. In the ``Supplement 1: Noether Symmetry Approach; Sect. (\ref{mathematical supplement})'' it is demonstrated that in both cases, $g(T)$ has the common form as $g_{0}\sqrt{-2(2m+1)T}$, while the forms of the other functions are different. Note that the aforementioned forms for $f(\varphi)$ are well-known and physical, therefore, a standard form (i.e. $g_{0}\sqrt{-2(2m+1)T}$ ) for $g(T)$ is used. It is important to mention that the forms $f(\varphi)=f_{0}\varphi^2$, and $g(T)=\sqrt{-6T}$ lead to late-time-accelerated expansion \cite{beh-intervention}. Besides Noether symmetry approach, this selection has other reasons: For a suitable form of the actions of \(F(T)\) gravity like
\begin{align*}
S=\int d^4x \; e \left[T+g(T) \right],
\end{align*}
it has been discussed in ref. \cite{J1} that a power-law form of correction term $g(T)\sim T^n \; (n>1)$ such as $T^2$ may remove the finite-time future singularity. Obviously, the correction term can be regarded as a cosmological constant when $n=0$. The model with $n=1/2$ may be helpful in realizing power-law inflation, and also describes little-rip and pseudo-rip cosmology \cite{J1}. Moreover, this form has correspondence with the cosmological constant Equation of State (EoS) in \(F(T)\) gravity \cite{ft1}. Due to these further reasons, choosing $g(T)\sim \sqrt{-T}$ is completely reasonable.

One advantageous selection for $U(\varphi, \varphi_{,\mu} \varphi^{,\mu})$ by assuming that two main parts of $U$ are separable is $h(\varphi) \dot{\varphi}$ where $h(\varphi)$ is an unknown function of the scalar field. However, as mentioned in ref. \cite{beh-intervention}, there are some reasons for this choice. Moreover, if the form of $U$ is not specified at first, then the problem poses a very difficult case. In this paper, we also use this clever choice. In order to have a better understanding of the action (\ref{action}) and these discussions, we may refer the readers to study ref. \cite{beh-intervention}. Therefore, by utilizing the Lagrange's method of undetermined coefficients, the action (\ref{action}) may be written as
\begin{align}\label{maction}
S = \int d^4x \; e \bigg[ f(\varphi)T - h(\varphi)\dot{\varphi} g(T) -\lambda \left(T+2\left(2m+1\right) \frac{\dot{B}^2}{B^2} \right)+ \frac{\omega(\varphi)}{2}\varphi_{,\mu} \varphi^{,\mu} - V(\varphi) \bigg],
\end{align}
where the Lagrange multiplier $\lambda$ is derived by varying the action (\ref{maction}) with respect to $T$
\begin{align}\label{lagrangian multiplier}
\lambda=f-h\dot{\varphi}g^{\tau}
\end{align}
in which the $\tau$ denotes a differentiation with respect to the torsion $T$. Thus, the point-like Lagrangian corresponding to the action (\ref{action}) becomes
\begin{align}\label{point like lagrangian}
L= fTB^{m+2}-h\dot{\varphi}gB^{m+2} -(f-h\dot{\varphi}g^{\tau})
(TB^{m+2}+2(2m+1)\dot{B}^2B^{m}) +\frac{1}{2}\omega\dot{\varphi}^2B^{m+2}
-VB^{m+2}.
\end{align}
The Euler-Lagrange equations for a dynamical system are
\begin{align}\label{euler2}
\frac{\partial L}{\partial q_{i}}- \frac{\mathrm{d}}{\mathrm{d}t}\left( \frac{\partial L}{\partial \dot{q}_{i}}\right)=0,
\end{align}
where $q_{i}$ are the generalized positions in the corresponding configuration space (i.e. $Q=\{q_{i}\}$). The energy function associated with the Lagrangian is given by
\begin{align}\label{euler3}
E_{L}=\sum_{i}\dot{q}_{i}\frac{\partial L}{\partial \dot{q}_{i}}-L.
\end{align}
Hence, according to the point-like Lagrangian (\ref{point like lagrangian}), the corresponding Euler-Lagrange equation for the scale factor $B$ would be
\begin{align}\label{fe1}
&4(2m+1)\left(f-h\dot{\varphi}g^{\tau}\right)\left(\frac{\ddot{B}}{B}\right)
\nonumber \\&+2m(2m+1)\left(f-h\dot{\varphi}g^{\tau}\right)
\left(\frac{\dot{B}}{B}\right)^2 \nonumber \\&+4(2m+1)\left(f^{\prime}\dot{\varphi}
-hg^{\tau}\ddot{\varphi}
-hg^{\tau\tau}\dot{\varphi}\dot{T}-h^{\prime}g^{\tau}
\dot{\varphi}^2\right)\left(\frac{\dot{B}}{B}\right)
\nonumber\\&+(m+2)(Tg^{\tau}-g)h\dot{\varphi}+\frac{1}{2}(m+2)\omega \dot{\varphi}^2-(m+2)V=0,
\end{align}
where the prime indicates a derivative with respect to the scalar field $\varphi$. For the scalar field, $\varphi$, the Euler-Lagrange equation takes the following form
\begin{align}\label{fe2}
&4(2m+1)hg^{\tau}\left(\frac{\ddot{B}}{B}\right)
\left(\frac{\dot{B}}{B}\right)+2m(2m+1)hg^{\tau}
\left(\frac{\dot{B}}{B}\right)^3 \nonumber \\&+
2(2m+1)\left(f^{\prime}+hg^{\tau \tau}\dot{T}\right)\left(\frac{\dot{B}}{B}\right)^2 \nonumber \\ &+(m+2)\left(\omega \dot{\varphi}+hg^{\tau}T-hg\right)\left(\frac{\dot{B}}{B}\right) \nonumber \\
&+hg^{\tau}T\dot{T}+\omega \ddot{\varphi}+\frac{1}{2}\omega^{\prime}\dot{\varphi}^2+V^{\prime}=0,
\end{align}
which is the Klein-Gordon equation. The energy function which is the $\binom{0}{0}$-Einstein equation, associated with the point-like Lagrangian (\ref{point like lagrangian}) is found as
\begin{align}\label{fe4}
2(2m+1)\left(2hg^{\tau}\dot{\varphi}-f\right)
\left(\frac{\dot{B}}{B}\right)^2+\frac{1}{2}\omega \dot{\varphi}^2+V=0.
\end{align}
And, the Euler-Lagrange equation for the torsion scalar $T$ reads
\begin{align}\label{fe3}
hg^{\tau\tau}\dot{\varphi}\left(T+2(2m+1)\frac{\dot{B}^2}{B^2}\right)=0.
\end{align}

\noindent\hrmybox{}{\section{Nature of the Scalar Potentials via the Reconstruction Method\label{Section05}}}\vspace{5mm}

In this section, finding a general formula for investigating the nature of the scalar potentials which works in at least some well-known cases is our objective. For this purpose, we utilize the extended reconstruction method proposed by A.Y. Kamenshchik, A. Tronconi, and G. Venturi \cite{ref1}. As mentioned above, first, for a physical and reasonable choice of $g(T)$ (i.e. $g_{0}\sqrt{-2(2m+1)T}$), we formulate a general formalism for considering the nature of the field potential. Substituting this form of $g(T)$, eqs. (\ref{fe1})-(\ref{fe4}) turn out to be
\begin{align}
&4(2m+1)f \left(\frac{\ddot{B}}{B}\right)+2m(2m+1)f
\left(\frac{\dot{B}}{B}\right)^2 \nonumber \\& +4(2m+1)
f^{\prime}\dot{\varphi} \left(\frac{\dot{B}}{B}\right)+2g_{0}(2m+1)h \ddot{\varphi} \nonumber
\\&+2g_{0}(2m+1)h^{\prime}\dot{\varphi}^2
+\left(\frac{1}{2}m+1\right)\omega \dot{\varphi}^2-(m+2)V=0, \label{rfe1} \\
&2g_{0}(2m+1)h\left(\frac{\ddot{B}}{B}\right)
+2g_{0}(2m+1)(m+1)h\left(\frac{\dot{B}}{B} \right)^2 \nonumber \\& -2(2m+1)f^{\prime} \left(\frac{\dot{B}}{B}\right)^2-(m+2)\omega \dot{\varphi}\left(\frac{\dot{B}}{B}\right) \nonumber \\&-\omega \ddot{\varphi}-\frac{1}{2}\omega^{\prime}\dot{\varphi}^2-V^{\prime}=0,
\label{rfe2} \\
&-2(2m+1)f B^m \dot{B}^2 -2(2mg_{0}+1)hB^{m+1}\dot{\varphi}\dot{B}\nonumber \\ &+\frac{1}{2}\omega B^{m+2} \dot{\varphi}^2+B^{m+2}V=0,\label{rfe4}
\end{align}
respectively. From eq. (\ref{fe3}), there are three possibilities: $i.$ $g^{\tau \tau}=0$ which does not hold for the taken form of $g$; $ii.$ $\dot{\varphi}=0$ which leads to a constant scalar field, hence, not only it is not suitable, but also is wrong when we apply this reconstruction approach (In what follows we show this); and $iii.$ the possibility of $T=-2(2m+1)\dot{B}^2/B^2$ which is the definition of the scalar torsion for LRS BI background under the assumption $A=B^m$ (See eq. (\ref{torsion1})). Therefore, eq. (\ref{fe3}) is satisfied automatically because of the last option. Hence, we leave it.\\
Eq. (\ref{rfe4}) yields
\begin{align}\label{v08}
V(\varphi)= 2(2m+1)f\left(\frac{\dot{B}}{B}\right)^2 +2(2mg_{0}+1)h\dot{\varphi}
\left(\frac{\dot{B}}{B}\right) -\frac{1}{2}\omega \dot{\varphi}^2.
\end{align}
It will be more convenient to consider all the functions as functions of the scale factor $B$ and not of time $t$. Correspondingly, eq. (\ref{v08}) may be rewritten as
\begin{align}\label{v8}
V(B)=H^2_{2}\bigg[2(2m+1)f &+2(2mg_{0}+1)hB\varphi_{,B} -\frac{1}{2}\omega B^2 \varphi^2_{,B}\bigg].
\end{align}
Note that the notations $X_{,y}\equiv \mathrm{d}X /\mathrm{d}y$ and also $X_{,yy}\equiv \mathrm{d}^2 X /\mathrm{d}y^2$ are used in this paper. The Klein-Gordon equation, (\ref{rfe2}), may be written down as
\begin{align}\label{v9}
&2(2mg_{0}+1)h\ddot{\varphi}+(m+2)\omega\dot{\varphi}^2
+2(2mg_{0}+1)h^{\prime}\dot{\varphi}^2 \nonumber \\&+2(2m+1)
\left(2f^{\prime}-g_{0}(m+2)h\right)H_{2}
\dot{\varphi} \nonumber \\& +4(2m+1)f\dot{H}_{2}=0,
\end{align}
in which the derivative of the scalar potential (i.e. $V^{\prime}=\mathrm{d}V/\mathrm{d}\varphi$) has been eliminated through the following relation which may be found by using eq. (\ref{v08}):
\begin{align}\label{v09}
V^{\prime}=\frac{\mathrm{d}V(\varphi)}{\mathrm{d}\varphi}&=2(2mg_{0}+1)
hH_{2}\frac{\ddot{\varphi}}{\dot{\varphi}}
-\omega \ddot{\varphi}-\frac{1}{2}\omega^{\prime}\dot{\varphi}^2 \nonumber \\& +2(2mg_{0}+1)H_{2}h^{\prime}\dot{\varphi}+4(2m+1)fH_{2} \frac{\dot{H}_{2}}{\dot{\varphi}}\nonumber \\& +2(2mg_{0}+1)h\dot{H}_{2}+2(2m+1)H^2_{2} f^{\prime}.
\end{align}
For a general case, it is more appropriate to use the dependence of the scalar field $\varphi$ upon the cosmological radius $B$. Therefore, eq. (\ref{v9}) should be rewritten as
\begin{align}\label{v11}
&2(2mg_{0}+1)h\varphi_{,BB}+2(2mg_{0}+1)
h^{\prime}\varphi^2_{,B}+(m+2)\omega \varphi^2_{,B} \nonumber \\&+2(2m+1)
\left(2f^{\prime}-g_{0}(m+2)h\right)\frac{\varphi_{,B}}{B}
\nonumber \\ &+2(2mg_{0}+1)h\varphi_{,B}\frac{H_{2,B}}{H_{2}}
+4(2m+1)\frac{f}{B}\frac{H_{2,B}}{H_{2}}=0.
\end{align}
Introducing the variable $\chi=\chi(B)$ which we define it as
\begin{align}\label{vg1}
\chi=\frac{\varphi_{,B}}{h},
\end{align}
eq. (\ref{v11}) turns out to be
\begin{align}\label{v14}
&\left(\frac{(m+2)\omega+4(2mg_{0}+1)h^{\prime}}{2(2mg_{0}+1)} \right)\chi^2
\nonumber \\&+ \frac{(2m+1)}{(2mg_{0}+1)}\left(2\frac{f^{\prime}}{h}
-g_{0}(m+2)\right)\frac{\chi}{B} \nonumber \\ &+\frac{2(2m+1)}{(2mg_{0}+1)}\frac{f}{h^2 B}\frac{H_{2,B}}{H_{2}}+\chi \frac{H_{2,B}}{H_{2}}+\chi_{,B}=0.
\end{align}
It is worth mentioning that choosing a true transformation is the most important part of this procedure and it is completely model-dependent. However, it is impossible to find a general transformation leading to analytically solvable general basic equation for all cases of coupling functions, but one may find one that works at least in several cases.\\
Now, let us reexpress eq. (\ref{v14}) by assuming a new function $M=M(B)$ as
\begin{align}\label{vg2}
\chi=\left( \frac{2(2mg_{0}+1)}{(m+2)\omega+4(2mg_{0}+1)h^{\prime}}\right)\frac{M_{,B}}{M}.
\end{align}
Therefore, we obtain the following basic general formalism
\begin{align}\label{v15}
&\left(\frac{2(2mg_{0}+1)}{(m+2)\omega+4(2mg_{0}+1)h^{\prime}}\right)\left( \frac{M_{,BB}}{M}
+\frac{M_{,B}}{M}\frac{H_{2,B}}{H_{2}}\right) \nonumber \\
&-2(2mg_{0}+1)\varphi_{,B}\left(\frac{(m+2)\omega^{\prime}+4(2mg_{0}+1)h^{\prime \prime}}{\left((m+2)\omega+4(2mg_{0}+1)h^{\prime}\right)^2}\right)
\nonumber \\
& \times \left(\frac{M_{,B}}{M}\right)+\left(\frac{2(2m+1)}{(m+2)\omega
+4(2mg_{0}+1)h^{\prime}}\right) \nonumber \\ & \times \left(2\frac{f^{\prime}}{h}-g_{0}(m+2) \right)\frac{1}{B}\frac{M_{,B}}{M} \nonumber \\
&+\frac{2(2m+1)}{(2mg_{0}+1)}\frac{f}{h^2 B}\frac{H_{2,B}}{H_{2}}=0.
\end{align}
It is important to notice from (\ref{vg1}) and (\ref{vg2}) that
\begin{align}\label{vg12}
M=\left(\frac{M_{0}}{c^2_{0}}\right)h^2\exp \left[\frac{(m+2)}{2(2mg_{0}+1)}
\int \frac{\omega}{h}d\varphi \right],
\end{align}
where $c_{0}$ and $M_{0}$ are the nonzero constants of integration. In section (\ref{examples1234}), we examine our obtained formulas, (\ref{v15}) and (\ref{vg12}), by considering some well-known cases and show that they work fairly well.\\

\noindent\hrmybox{}{\section{Considering some well-known cases\label{examples1234}}}\vspace{5mm}

The obtained basic equation is used in this section to investigate two well-known coupling cases ($f=f_{0}\varphi^2$ and $f=f_{0}$) by taking three different matter contents (`Barotropic fluid', `Cosmological constant', and `Modified Chaplygin gas') into account.\\

\noindent\rmybox{}{\subsection{The case $f(\varphi)=f_{0}\varphi^2$}}\vspace{5mm}

In this sub-section, we study the most important coupling function $f(\varphi)=f_{0}\varphi^2$. As mentioned earlier, the form of other unknown coupling functions are fixed by the Noether symmetry approach. According to it (See eq. (\ref{sol for NS2})), we have:
\begin{align}\label{res sol for NS2}
f(\varphi)=f_{0}\varphi^2, \quad \omega(\varphi)=\omega_{0}, \quad h(\varphi)=h_{0}\varphi,
\end{align}
where, without loss of generality (only for convenience), we have set $k_{1}=1$.
Therefore, the basic equation (i.e. Eq. (\ref{v15})) takes the following form
\begin{align}\label{v15-1}
\lambda_{1}\left( \frac{M_{,BB}}{M}
+\frac{M_{,B}}{M}\frac{H_{2,B}}{H_{2}}\right)
+\lambda_{2}\frac{1}{B}\frac{M_{,B}}{M}
+\lambda_{3}\frac{1}{B}\frac{H_{2,B}}{H_{2}}=0,
\end{align}
where
\begin{align}\label{Ayandeh}
\lambda_{1}&=\frac{2(2mg_{0}+1)}{(m+2)\omega_{0}+4(2mg_{0}+1)h_{0}},\nonumber \\ \lambda_{2}&=\left(\frac{2(2m+1)}{(m+2)\omega_{0}+4(2mg_{0}+1)h_{0}}\right) \left(4\frac{f_{0}}{h_{0}}-g_{0}(m+2) \right), \nonumber \\ \lambda_{3}&=\frac{2(2m+1)}{(2mg_{0}+1)}\frac{f_{0}}{h^2_{0} }.
\end{align}
Limpidly, if one puts the form of $H_{2,B}/H_{2}$ in (\ref{v15-1}), then it may be solved. In what follows, the evolution of the Hubble parameter is inserted by three different matter contents.

On the other hand, from eq. (\ref{vg12}) one has
\begin{align}\label{shekle M}
\varphi=C M^{\frac{2h_{0}(2mg_{0}+1)}{4h_{0}(2mg_{0}+1)+\omega_{0}(m+2)}},
\end{align}
in which all the constants of integration have been absorbed in $C$. Therefore, the form of the scalar field, $\varphi$, is specified if we identify the form of $M(B)$. It is beneficial to define
\begin{align}\label{theta}
\theta \equiv \frac{4h_{0}(2mg_{0}+1)+\omega_{0}(m+2)}{2h_{0}(2mg_{0}+1)},
\end{align}
because this term appears very much in this paper.

As mentioned earlier, the analysis of this paper would be depended on the ``Class-1'' of ref. \cite{hashem}. Hence, according to the ``Table-I'' in ref. \cite{hashem}, the coupling functions, $f(\varphi)=f_{0}\varphi^2$, and $h(\varphi)=h_{0}\varphi$, which give the conditions $\mathfrak{B}[\varphi,0;f(\varphi)]=2 \geq 0$ and $\mathfrak{B}[\varphi,0;h(\varphi)]=1 \geq 0$, are of decreasing coupling functions with time, `Type-I' (i.e. They always decrease with time). Therefore, four applicable other conditions, namely
\begin{align}
\mathfrak{B}[A,\gamma_{01};f(A)]\leq 0, \quad \mathfrak{B}[B,\gamma_{02};f(B)]\leq 0, \label{chaharshart1}\\
\mathfrak{B}[A,\gamma_{03};h(A)]\leq 0, \quad \mathfrak{B}[B,\gamma_{04};h(B)]\leq 0,\label{chaharshart2}
\end{align}
must also be held. In the following examples, we are able to consider them after obtaining $\varphi(B)$ and $\varphi(A)$.\\

\noindent\Rrmybox{}{\subsubsection{\textbf{Barotropic fluid}}}\vspace{5mm}

A particular case of the perfect fluid is the barotropic fluid which its density is a function of pressure only, viz.,
\begin{align}\label{vgb1}
P=\Upsilon \rho; \qquad 0<\Upsilon<1,
\end{align}
where $P$ and $\rho$ are pressure and density, respectively, while $\Upsilon$ is the EoS parameter. Using (\ref{vgb1}) and the conservation equation (\ref{chegali}), one obtains
\begin{align}\label{vgb4}
\rho=\rho_{0} B^{-(m+2)(1+\Upsilon)},
\end{align}
where $\rho_{0}$ is a positive integration constant.
Inserting (\ref{vgb1}) into the right side of eq. (\ref{hard1}) and combining with (\ref{hard2}) we get
\begin{align}\label{unieq}
\left(\frac{m^2+m+4}{2}\right)\left(\frac{\rho}{2m+1}\right)
+\left(\frac{m+3}{2}\right)
\dot{H}_{2}= - \Upsilon \rho.
\end{align}
Substituting (\ref{vgb4}) into eq. (\ref{unieq}) and solving it (by using $\dot{H}_{2}=BH_{2}H_{2,B}$) leads to
\begin{align}\label{v18}
H_{2}(B)=H_{0b} \; H_{0b1} \; B^{\frac{-(1+\Upsilon)(m+2)}{2}},
\end{align}
where $H_{0b}$ is an integration constant which must be a positive real number due to the expanding nature of the universe ($\mathfrak{B}[t,0;B(t)]>0$), and
\begin{align}\label{h0b1}
H_{0b1}=\sqrt{\left(\frac{m^2+m+4}{1+2m}+2\Upsilon\right)
\frac{2\rho_{0}}{(1+\Upsilon)(m+2)(m+3)}}.
\end{align}
Therefore
\begin{align}\label{vgn1}
\frac{H_{2,B}}{H_{2}}=\frac{-(1+\Upsilon)(m+2)}{2B},
\end{align}
is obtained. This equation indicates the evolution of the Hubble parameter owing to the barotropic fluid. Consequently, the dimensionless function $\mathfrak{B}[a,0;H(a)]$ would be
\begin{align}
\mathfrak{B}[a,0;H(a)]=\frac{-3(1+\Upsilon)}{2}.
\end{align}
Obviously, we obtain: $-3<\mathfrak{B}[a,0;H(a)]<-1.5$. Therefore, according to ``Table III'' and ``Figure 1'' in ref. \cite{hashem}, it determines decelerated expansion and quintessence phase for this model which is consistent with the barotropic fluid. Note that $\Upsilon=1/3$ corresponds to the radiation-dominated era in which the universe was of decelerating nature (i.e. $q_{\mathrm{rad.}}>0$).\\
Note that however in ref. \cite{hashem} it was stated that all the conditions must be considered for each direction in anisotropic universes and hence, as a rule, we must consider $\mathfrak{B}\text{-function}$ for each direction (i.e. $\mathfrak{B}[A,0;H_{1}(A)]$ and $\mathfrak{B}[B,0;H_{2}(B)]$) and discuss its type of expansion. But, in this paper, it is needless to do it, because according to the following relations
\begin{align}\label{HAMID}
\mathfrak{B}[B,0;H_{2}(B)]=\left(\frac{m+2}{3}\right)
\mathfrak{B}[a,0;H(a)] ,\\
\mathfrak{B}[A,0;H_{1}(A)]=\left(\frac{m(m+2)}{3}\right)
\mathfrak{B}[a,0;H(a)] ,
\end{align}
and the factors $(m+2)/3$ and $m(m+2)/3$ which are positive and very close to $1$, and therefore they do not change the conditions mentioned in ref. \cite{hashem} for the type of expansion and phase, hence, it is sufficient to discuss with $\mathfrak{B}[a,0;H(a)]$ only. In other words, in our case in which we set the condition $A=B^{m}$, there is no difference that we work with which of $\mathfrak{B}[B,0;H_{2}(B)]$, $\mathfrak{B}[A,0;H_{1}(A)]$, and $\mathfrak{B}[a,0;H(a)]$ since they produce the same results for the status of the expansion and also for the type of the phase of the universe.\\

With eq. (\ref{vgn1}) at hand, by taking
\begin{align}\label{savabet}
s_{1}&=\frac{2(2mg_{0}+1)}{\omega_{0}(m+2)+4h_{0}(2mg_{0}+1)}, \nonumber \\
s_{2}&=\left(\frac{2(2mg_{0}+1)}{\omega_{0}(m+2)+4h_{0}(2mg_{0}+1)}\right)
\left(\frac{-(1+\Upsilon)(m+2)}{2}\right)
\nonumber \\&+\left(\frac{2(2m+1)}{\omega_{0}(m+2)+4h_{0}(2mg_{0}+1)} \right)
\left(4\frac{f_{0}}{h_{0}}-g_{0}(m+2)\right), \nonumber \\
s_{3}&=\frac{-f_{0}(2m+1)(1+\Upsilon)(m+2)}{h^2_{0}(2mg_{0}+1)},
\end{align}
eq. (\ref{v15-1}) would be
\begin{align}\label{kholaseh}
s_{1}\frac{M_{,BB}}{M}+\frac{s_{2}}{B}\frac{M_{,B}}{M}
+\frac{s_{3}}{B^2}=0,
\end{align}
whose solution is
\begin{align}\label{sol for khol}
M(B)=c_{1}B^{p_{1}}
+c_{2}B^{p_{2}},
\end{align}
where $c_{1}$ and $c_{2}$ are integration constants, and $p_{1}$ and $p_{2}$ are
\begin{align}\label{p}
p_{1}&=\frac{{s_{1}-s_{2}+\sqrt{s^2_{1}-2s_{1}s_{2}-4s_{1}s_{3}+s^2_{2}}}}{2s_{1}}, \nonumber \\ p_{2}&=\frac{{s_{1}-s_{2}-\sqrt{s^2_{1}-2s_{1}s_{2}-4s_{1}s_{3}+s^2_{2}}}}{2s_{1}}.
\end{align}
Therefore, considering eq. (\ref{shekle M}), the corresponding scalar field would be
\begin{align}\label{csf1}
\varphi(B)
=C\left(c_{1}B^{p_{1}}+c_{2}B^{p_{2}} \right)^{\frac{1}{\theta}}.
\end{align}
Since in general, $p_{1}\neq p_{2}$, $p_{1} \neq 0$, and $p_{2} \neq 0$, hence we cannot invert this equation to reach the explicit form of the scale factor. It imposes that we consider two cases: $i$. $c_{1}=0$, and $ii$. $c_{2}=0$. They lead to
\begin{align}\label{nvg231}
B(\varphi)=\left(\frac{\varphi}{Cc_{1,2}}\right)^{\frac{\theta}{p_{1,2}}} \quad \Longrightarrow \quad
A(\varphi)=\left(\frac{\varphi}{Cc_{1,2}}\right)^{\frac{m\theta}{p_{1,2}}}.
\end{align}
Note that the subscripts $1$ and $2$ correspond to $c_{2}=0$ and $c_{1}=0$, respectively. As is observed, the scale factors have the power law natures.
According to the $\mathfrak{B}\text{-function}$ method, the conditions $\mathfrak{B}[t,0;A(t)]>0$ and $\mathfrak{B}[t,0;B(t)]>0$ imply
\begin{align}\label{nvg2311.1}
\frac{\theta}{p_{1,2}}=\frac{4h_{0}(2mg_{0}+1)
+\omega_{0}(m+2)}{2h_{0}p_{1,2}(2mg_{0}+1)} < 0,
\end{align}
and
\begin{align}\label{nvg2311.2}
\frac{m\theta}{p_{1,2}}=\frac{4h_{0}m(2mg_{0}+1)
+\omega_{0}m(m+2)}{2h_{0}p_{1,2}(2mg_{0}+1)} < 0.
\end{align}
Limpidly, these inequalities compel a condition: $m>0$. The signs of $Cc_{1}$ and $Cc_{2}$ are not important at all, because they demonstrate the directions of expanding. The positive signs of the Hubble parameters indicate an expanding universe, not scale factors. The conditions (\ref{chaharshart1})-(\ref{chaharshart2}) give $p_{1,2}/\theta \leq 0$ and $mp_{1,2}/\theta \leq 0$. But, since the common range of all conditions must be established, hence, pursuant to (\ref{nvg2311.1})-(\ref{nvg2311.2}), $p_{1,2}/\theta \leq 0$ and $mp_{1,2}/\theta \leq 0$ reduce to $p_{1,2}/\theta < 0$ and $mp_{1,2}/\theta < 0$, respectively.

It is worth noting that the condition (\ref{nvg2311.1}) may be achieved via another way as well:\\
For the radiation-dominated era, $\Upsilon=1/3$, the average scale factor is $a_{ave.} \simeq t^{2/3}$. If the scale factors obtained here, (\ref{nvg231}), are compared with it, yield
\begin{align}\label{*780}
\left(\frac{\varphi}{Cc_{1,2}} \right)^{\frac{\theta}{p_{1,2}}\frac{(m+2)}{3}}\simeq t^{\frac{2}{3}}
\quad \Longrightarrow \quad\varphi=Cc_{1,2} \; t^{\frac{2}{(m+2)}\frac{p_{1,2}}{\theta}}
\end{align}
where we have used $a_{ave.}=B^{(m+2)/3}$. Therefore, in view of $\mathfrak{B}[t,0;\varphi(t)]<0$, we obtain $(p_{1,2} / \theta)<0$.

Substituting eqs. (\ref{res sol for NS2}), (\ref{v18}), and (\ref{csf1}), into (\ref{v8}), the scalar field potentials turn out to be
\begin{align}\label{v24}
V(B)_{1,2}=V_{01,02}
B^{-(1+\Upsilon)(m+2)+\frac{2p_{1,2}}{\theta}}
\end{align}
where the subscripts $01$ and $02$ correspond to the cases 1 and 2, respectively and
\begin{align}\label{Ev24}
V_{01,02}=&H_{0b}\left(\frac{m^2+m+4}{1+2m}+2\Upsilon\right)
\nonumber \\ & \times \left(\frac{2\rho_{0}C^2 c^2_{1,2}}{(1+\Upsilon)(m+2)(m+3)}\right)
\bigg[2f_{0}(2m+1) \nonumber \\ &+2h_{0}p_{1,2}(2mg_{0}+1)\theta^{-1}
-\frac{1}{2}\omega_{0}p^2_{1,2} \theta^{-2} \bigg].
\end{align}
If we set
\begin{align}\label{shart1}
-(1+\Upsilon)(m+2)+\frac{2p_{1,2}}{\theta} < 0,
\end{align}
which is compatible with the condition $\mathfrak{B}[B, \gamma_{0}=0; V(B)] \leq 0$, then according to the expanding nature of the scale factor, the behavior of the potential versus the scale factor will be physical. Also, it may easily be shown that we must take the positive ranges of $m$ if we want to have physically acceptable nature for the potential.

Writing the potentials (\ref{v24}) in terms of the scalar field by using eq. (\ref{nvg231}) gives
\begin{align}\label{VNS}
V(\varphi)_{1,2}=V_{0}\varphi^{-(1+\Upsilon)(m+2)p_{1,2} \theta^{-1}+2}
=V_{0}\varphi^{\mu_{1,2}+2},
\end{align}
in which all the coefficients have been absorbed in $V_{0}$. The form of the potential obtained from the Noether symmetry approach (i.e. $V=V_{0}\varphi^{-2}$) is recovered by putting
\begin{align}\label{cNS2}
\mu_{1,2} = \frac{-(1+\Upsilon)(m+2)p_{1,2}}{\theta}= -4.
\end{align}
But, establishing this condition is wrong, since according to the aforementioned conditions (i.e. (\ref{nvg2311.1})--(\ref{nvg2311.2}), $m>0$, and $0<\Upsilon<1$), $\mu_{1,2}$ cannot be negative. Therefore the potential emerged by requiring the existence of  Noether symmetry does not hold here. This paradox occurs in some cases (for example, see \cite{capp1,capp2,capp3,capp4}). Setting $\mu_{1,2}=2N-2$, where $N=2, 3, ...$, the well-known potentials (i.e. $V=V_{0}\varphi^4, \varphi^6, \varphi^8, ...$) are obtained. Note that
\begin{align*}
\frac{2h_{0}p_{1,2}(2mg_{0}+1)} {4h_{0}(2mg_{0}+1)+\omega_{0}(m+2)}=\frac{p_{1,2}}{\theta}
\end{align*}
cannot be zero and infinity (i.e. $p_{1,2}\neq 0, \infty$; $\theta \neq 0, \infty$; $g_{0} \neq -1/2m$; and $\cdots$), otherwise the form of the scale factors would be undetermined. Therefore, we conclude that $\mu_{1,2} > 0$. Nonetheless, if we restrict ourselves to a strong condition as $0<\mu_{1,2} \lll 2$, then the quadratic form of the potential (i.e. $V=V_{0}\varphi^2$) is procured. Taking positive values of $\mu$, $\mathfrak{B}[\varphi, \phi_{0}=0; V(\varphi)] \geq 0$ is satisfied as well, because of $\varphi V^{\prime}_{1,2}/V_{1,2}=\mu_{1,2} +2\geq 0$. Note that, however, it allows $\mu_{1,2} \geq -2$, but the common domains of the conditions must be adopted, therefore it reduces to $\mu_{1,2}>0$. Clearly, the type of the potential (\ref{VNS}) is of Decreasing Scalar Potential with respect to time (`D.S.P.'), and therefore it can never be of Increasing Scalar Potential with respect to time (`I.S.P.') when we reconstruct with `Class-1'.\\

Pursuant to the function \cite{tracker,hashem}
\begin{equation}\label{take2}
\Gamma=\frac{VV^{\prime \prime}}{\left(V^{\prime}\right)^2}
=\frac{\mathfrak{B}[\varphi,0;V^{\prime}(\varphi)]}
{\mathfrak{B}[\varphi,0;V(\varphi)]},
\end{equation}
where $V=V(\varphi)$, three different solutions are possible for any form of the potentials:
\begin{equation}\label{take2}
\left\{
  \begin{array}{ll}
    \Gamma<1, & \hbox{Thawing;} \nonumber \\
    \Gamma=1, & \hbox{Scaling;} \nonumber \\
    \Gamma>1, & \hbox{Tracker.}
  \end{array}
\right.
\end{equation}
Therefore, the potentials (\ref{VNS}) are of `Thawing' type, because we found that $\mu_{1,2}>0$, hence
\begin{equation}\label{track2}
\Gamma=\frac{\mu_{1,2}+1}{\mu_{1,2}+2}<1.
\end{equation}
It is clearly observed that the there is no evolution for $\Gamma$ hence it causes retaining the `Thawing' type for the potential.\\

\noindent\Rrmybox{}{\subsubsection{\textbf{Cosmological constant}}}\vspace{5mm}

Accelerating picture of the expanding nature of the universe requires a \textit{negative pressure} together with a \textit{positive density}. The `cosmological constant' is one of the candidates proposed to this purpose. In the case of a cosmological constant, we have $P=-\rho$. Inserting this state equation into the conservation equation (Eq. (\ref{chegali})) leads to a constant energy density (i.e. $\dot{\rho}=0 \longrightarrow \rho=\rho_{0}>0$).\\

\noindent\qrmybox{red}{\paragraph{\textbf{The first case: LRS Bianchi I ($A=B^m; \; m\neq 0, 1$):}}}\\

For the case $m\neq0,1$, after performing the similar calculations to the `Barotropic fluid' case, one may obtain
\begin{align}\label{v26-1}
H_{2}=H_{0c}\sqrt{\frac{2\rho_{0}}{m+3}\left(2-\frac{m^2+m+4}{2m+1}\right)
\ln\left(B\right)},
\end{align}
where $H_{0c}$ is an integration constant which must be a positive real number, for holding the expansion nature of the universe, and also one of the options $\{(m>2 \; \text{ or } \; 0<m<1) \quad \& \quad 0<B<1\}$ or $\{1<m<2 \quad \& \quad B>1 \}$ must be justified, for having a real-valued Hubble parameter. According to (\ref{m-range}), the former one is completely nonphysical when $m>2$, for it is beyond the observational limits, and almost physical when $0<m<1$ that in view of (\ref{m-range}) must reduce to $N^{-}_{0} \leq m<1$; and the latter one is somewhat larger than observational limits, so the physics of the problem implies that we use a stronger bound as $1 < m \leq N^{+}_{0}$ instead of $1<m<2$. However, in what follows we show that the first option cannot be true for some reasons and one must pick up $1 < m \leq N^{+}_{0}$.\\
Pursuant to (\ref{v26-1}), we obtain
\begin{align}\label{v26-2}
\frac{H_{2,B}}{H_{2}}=\frac{1}{2B\ln(B)}
\end{align}
for corresponding directional Hubble parameters and its evolution. Therefore, the function $\mathfrak{B}[a,0;H(a)]$ turns out to be
\begin{align}\label{09047}
\mathfrak{B}[a,0;H(a)]=\frac{3}{2(m+2)}\frac{1}{\ln(B)}.
\end{align}
Hence, we obtain Table (\ref{00table1-dec1}) in which we have set $\vartheta_{3}=-2(m+2)/3$.\\
The reason for depending the status of expansion of the universe (Accelerated/Decelerated) upon the scale factor $B$ is anisotropy property (i.e. $m\neq 1$) of the background studied (Note that the FRW space-time geometry is treated separately in what follows). As mentioned earlier, the common domain of all conditions must be considered, therefore, in this stage one cannot state that which of the conditions in Table (\ref{00table1-dec1}) is satisfied.\\
By substituting the formula (\ref{v26-2}) into (\ref{v15-1}) we obtain
\begin{align}\label{v28.1}
l_{1}\left(\frac{M_{,BB}}{M}+\frac{1}{2B\ln(B)}\frac{M_{,B}}{M}\right)
+\frac{l_{2}}{B}\frac{M_{,B}}{M}+\frac{l_{3}}{B^2\ln(B)}=0,
\end{align}
where
\begin{align}
l_{1}&=\frac{2(2mg_{0}+1)}{\omega_{0}(m+2)+4h_{0}(2mg_{0}+1)},\label{v28-1.1} \\
l_{2}&=\frac{2(2m+1)}{\omega_{0}(m+2)+4h_{0}(2mg_{0}+1)}
\left(4\frac{f_{0}}{h_{0}}-g_{0}(m+2)\right),\label{v28-1.2}\\
l_{3}&=\frac{f_{0}(2m+1)}{h^2_{0}(2mg_{0}+1)}.\label{v28-1.3}
\end{align}
Solving this equation without any assumption for the coefficients, $l_{1}$, $l_{2}$, and $l_{3}$, is challenging, since it leads to a very complicated integral in terms of $\mathbf{Whittaker}$ functions. But there are two ways for achieving a suitable solution: 1- Letting the coefficients to be special constants, such as $l_{1}=l_{2}=1$ and $l_{3}=-1/4$, and 2- Assuming $l_{1} \neq l_{2}$ and multiplying both sides of the equation by $M(B)$.\\
\begin{table*}
\caption{Different behaviors of expansion of the universe and their corresponding amounts of the scale factors in cosmological constant case} \label{00table1-dec1}
\centering
\begin{tabular}{l r}
\toprule[1.7pt]
\textbf{Conditions}&\textbf{The status of the universe}\\[0.5ex]
\toprule[1.4pt]
$\ln^{-1}(B)< \vartheta_{3}$
&\qquad Decelerated expansion and Quintessence phase\\[1.3ex]
\toprule[0.7pt]
$B=\exp(\vartheta_{3})$
&\qquad The inflection point and Quintessence phase\\[1.3ex]
\toprule[0.7pt]
$\vartheta_{3}<\ln^{-1}(B)<0$
& \qquad Accelerated expansion and Quintessence phase\\[1.3ex]
\toprule[0.7pt]
$\ln^{-1}(B)< -\vartheta_{3}$
& \qquad Super-accelerated expansion and Phantom phase\\[1.3ex]
\toprule[1.7pt]
\end{tabular}
\end{table*}

\noindent\rgoldmybox{red}{$\textcolor[rgb]{0.00,0.25,0.50}{\mathbf{\ast} \textbf{1- The First Way $l_{1}=l_{2}=1$, and $l_{3}=-1/4$:}}$}\\

Taking $l_{1}=l_{2}=1$, and $l_{3}=-1/4$, the solution would be:
\begin{align}\label{29.1}
M(B)=c_{3}\exp\left[-\sqrt{\ln(B)}\right]+c_{4}\exp\left[\sqrt{\ln(B)}\right],
\end{align}
where $c_{3}$ and $c_{4}$ are integration constants. Since $M(B)$ must be a real function of the scale factor $B$, therefore, in each time the scale factor $B$ is in the range $(1,+\infty)$. The real numbers of the range $[0,1)$ lead to phase (i.e. $M=c_{3}\exp[-ic]+c_{4}\exp[ic]$ where $c$ is a real constant), and $B=1$ gives $M(B)=constant$. However, the sign of the scale factor can be negative, but it is forbidden here because the real numbers of the range $(-\infty,0)$ lead to the complex numbers for logarithmic function.
Note that other assumptions for the values of $l_{1}$, $l_{2}$, and $l_{3}$ are also considerable (for example: $l_{1}=0$ leads to $M(B)=c_{5}\left(\ln(B) \right)^{-l_{3}/l_{2}}$, or another case as $l_{2}=l_{3}=0$ implies $M(B)=c_{6}+c_{7}\int\frac{dB}{\sqrt{\ln(B)}}$).
Because the arguments of the exponential functions are different, hence, we must choose one of the constants $c_{3}$ or $c_{4}$ equal to zero. In this case, we are able to invert the function (\ref{29.1}). It is worth mentioning that there are two other options: \textit{i}) $M(B)=\sinh[\sqrt{\ln(B)}]$ when one takes $c_{3}=-1/2$, $c_{4}=1/2$ and \textit{ii}) $M(B)=\cosh[\sqrt{\ln(B)}]$ when $c_{3}=c_{4}=1/2$; but both cases lead to uninteresting forms for the potential, hence we leave them. The $\mathfrak{B}\text{-function}$ method can help us to select the true sign of the argument of the exponential function. Hence, by keeping both, we take (\ref{29.1}) as
\begin{align}\label{ras1}
M(B)=c_{4,3}\exp\left[\pm\sqrt{\ln(B)}\right].
\end{align}
Therefore, according to (\ref{shekle M}), it yields
\begin{align}\label{moa-c-01}
\varphi(B)=C \left[c_{4,3} \exp \left(\pm\sqrt{\ln(B)}\right)\right]^{\frac{1}{\theta}},
\end{align}
and consequently
\begin{align}\label{moa-c-1}
B(\varphi)=\exp \left[\left(\ln \left(\frac{1}{c_{4,3}}\left[\frac{\varphi}{C} \right]^{\theta} \right)\right)^2 \right].
\end{align}
In view of $\mathfrak{B}[B,\alpha_{0}=0;\varphi(B)]<0$, and also the conditions (\ref{chaharshart1})-(\ref{chaharshart2}), both positive and negative signs in (\ref{ras1}) are admissible provided that $\theta$ be negative and positive, respectively:
\begin{equation}\label{N1}\begin{split}
\left\{
                                    \begin{array}{ll}
                                      \varphi=C \left[c_{4,3} \exp \left(-\sqrt{\ln(B)}\right)\right]
                                      ^{\frac{1}{\theta}} & \Longleftrightarrow \theta >0; \\ \\
                                      \varphi=C \left[c_{4,3} \exp \left(+\sqrt{\ln(B)}\right)\right]
                                      ^{\frac{1}{\theta}} & \Longleftrightarrow \theta <0.
                                    \end{array}
                                  \right.
\end{split}\end{equation}
Regarding $\mathfrak{B}[\varphi, \beta_{0}=0; B(\varphi)]<0$, we have
\begin{align}\label{shartbt1}
2\theta
\ln\left(\frac{1}{c_{4,3}}\left[\frac{\varphi}{C}\right]
^{\theta}\right)<0.
\end{align}
This physical condition is satisfied by two ways:
\begin{equation}\label{er1-shartbt1}\begin{split}
\left\{
  \begin{array}{ll}
    \text{1: }\; 0<\frac{1}{c_{4,3}}\left(\frac{\varphi}{C}\right)
^{\theta}<1, & \Longleftrightarrow \theta>0; \\ \\
    \text{2: }\; \frac{1}{c_{4,3}}\left(\frac{\varphi}{C}\right)
^{\theta}>1, & \Longleftrightarrow \theta<0.
  \end{array}
\right.
\end{split}\end{equation}
Clearly,
\begin{align}\label{e111}
c_{4,3}>0, \qquad \text{and} \qquad \frac{\varphi}{\left|\varphi \right|} \frac{C}{\left|C \right|}>0,
\end{align}
are emerged by both conditions of (\ref{er1-shartbt1}). Inserting (\ref{N1}) into (\ref{er1-shartbt1}) we obtain
\begin{equation}\label{N2}\begin{split}
\left\{
  \begin{array}{ll}
    1: \;  0<\exp\left(-\sqrt{\ln(B)}\right)<1, & \Longleftrightarrow \theta>0; \\ \\
    2: \;  \exp\left(+\sqrt{\ln(B)}\right)>1, & \Longleftrightarrow \theta<0.
  \end{array}
\right.
\end{split}\end{equation}
Because we have set $B >1$, hence the left parts of the above conditions are automatically satisfied. By the same way, for another scale factor, $A(\varphi)$, which defined as \(A(\varphi)= B^m(\varphi)\), we gain the same conditions and as well as one another: \(m>0\) (Note that the scale factor $A$ is not an independent scale factor because of the physical condition $A=B^m$, hence the conditions for the scale factor $A$ do not give further physical conditions than the scale factor $B$ excluding the condition $m>0$, and for this reason, from now on, we ignore to consider the conditions for $A$). Therefore, this model indicates super-accelerated expansion together with phantom phase only, because the last condition in Table (\ref{00table1-dec1}) is satisfied (Note that $m>0$ and in each time $B \in (1,+\infty)$). By comparing this with the options obtained from eqs. (\ref{v26-1}) and (\ref{m-range}), it may easily be found that in order to have a real-valued Hubble parameter, $m$ must be in $(1,N^{+}_{0}]$.\\
Because $l_{1}$ has here been set to $1$, so $\theta=h^{-1}_{0}$ (See eqs. (\ref{v28-1.1}) and (\ref{theta}); $l_{1}=(\theta h_{0})^{-1}$). Moreover, because we took $l_{3}=-1/4$ and $m>0$, thus $f_{0}(2mg_{0}+1)<0$ (See eq. (\ref{v28-1.3})). Therefore, with regards to the range of $m$ for this case, $(1,N^{+}_{0}]$, we learn that $f_{0}$ and $g_{0}$ are connected together via following conditions:
\begin{equation}
\left\{
  \begin{array}{ll}
   \text{1: } f_{0}<0 &\Longleftrightarrow g_{0}> \frac{-1}{2} \\ \\
   \text{2: } f_{0}>0 &\Longleftrightarrow g_{0}< \frac{-1}{2N^{+}_{0}}.
  \end{array}
\right.
\end{equation}
However, it seems at first sight that the amounts of these parameters are not important and in data analysis we can set them independently, but here we observe that it is not a true idea. The physical ranges of these constant parameters are tightly coupled. Moreover, finding the physical domains of the constant parameters is a time-consuming process, thus, these physical bounds make our work easy in plotting and in the data analysis.

Substituting eqs. (\ref{res sol for NS2}), (\ref{v26-1}), and (\ref{moa-c-01}), into (\ref{v8}), the scalar field potential in terms of the scale factor would be
\begin{align}\label{pot-cc-B}
V(B)=V_{occ} \left[\exp\left(\sqrt{\ln(B)}\right) \right]^{\frac{2}{\theta}}\ln(B),
\end{align}
where
\begin{align}\label{c-pot-cc-B}
V_{occ}=&\frac{2\rho_{0}H^2_{oc}C^2c^{\frac{2}{\theta}}_{4,3}}{m+3}
\left(2-\frac{m^2+m+4}{2m+1}\right) \nonumber \\ & \times \left[2(2m+1)f_{0}
+\frac{h_{0}(2mg_{0}+1)}{\theta}
-\frac{\omega_{0}}{8\theta^2}\right].
\end{align}
Pursuant to the ``Table I'' in ref. \cite{hashem}, one of the sets of the potential conditions must be held at a very short time interval. Here, we go through both sets of conditions to see which of them would be true.\\

$\bullet$ \textbf{The first set of the potential conditions (Type-I of the scalar functions: Decreasing scalar-field function w.r.t. time):}

We start with the condition $\mathfrak{B}[B, \gamma_{0}=0; V(B)] \leq 0$ which here reads
\begin{align}\label{sh-cc-V-B}
\frac{\sqrt{\ln(B)}+\theta}{\theta \ln(B)} \leq 0.
\end{align}
This condition restricts the parameters further than previous, for it is satisfied only for negative values of $\theta$:
\begin{equation}\label{N4}\begin{split}
\left\{
  \begin{array}{ll}
    \text{1- If } \theta>0 \; \Longrightarrow  &{} \frac{\sqrt{\ln(B)}+\theta}{\theta \ln(B)} \text{ is positive!}\divideontimes (\text{See } (\ref{sh-cc-V-B})), \\ &{} \text{ so $\theta>0$ is ineligible;} \\ \\
    \text{2- If } \theta<0 \; \Longrightarrow &{} \frac{\sqrt{\ln(B)}-|\theta|}{-|\theta| \ln(B)} \text{ can be negative},\\ &{} \text{ so it holds only for $B\geq \exp \left(\theta^2\right)$.}
  \end{array}
\right.
\end{split}\end{equation}
Therefore, despite $\mathfrak{B}[B, \alpha_{0}=0; \varphi(B)]<0$ which allows the existence of both signs in eq. (\ref{moa-c-01}), but it has here been found that $\mathfrak{B}[B, \gamma_{0}=0; V(B)] \leq 0$ accepts only the positive sign of the solution and also $B \in [\theta^2, \infty)$, hence we arrive at $\theta < -1$ (Because we first obtained that $B \in (1, \infty)$ and here we gained $\theta<0$ and $B \in [\theta^2, \infty)$).\\
Using eqs. (\ref{moa-c-1}) and (\ref{pot-cc-B}), the potential in terms of the scalar field turns out to be
\begin{align}\label{N5}
V(\varphi)=\tilde{V}_{0cc}
\left[\frac{\ln\left(c_{4,3}\left[\frac{C}{\varphi} \right]^\theta\right)}{\varphi} \right]^2,
\end{align}
where $\tilde{V}_{0cc}=V_{0cc}C^2c^{\frac{2}{\theta}}_{4,3}$. The Taylor series of this scalar potential function would be
\begin{align}\label{Taylor96}
V(\varphi)=\tilde{V}_{0cc}\sum_{j=0}^{\infty}f_{j}(\alpha, \theta)\left(\varphi - 1 \right)^{j},
\end{align}
where
\begin{align}\label{Taylor97}
&\alpha=c_{4,3}C^{\theta}, \nonumber \\
&f_{0}(\alpha, \theta)=\ln^2(\alpha), \qquad f_{1}(\alpha, \theta)=-2\ln^2(\alpha)-2 \theta \ln(\alpha), \nonumber \\
&f_{2}(\alpha, \theta)=3\ln^2(\alpha)+5 \theta \ln(\alpha)+\theta^2, \qquad \cdots.
\end{align}
The interesting feature of this series is that the constant and linear terms of the scalar field exist when $\alpha \neq 1$.
The condition $\mathfrak{B}[\varphi, \phi_{0}=0; V(\varphi)] \geq 0$ for this potential, (\ref{N5}), turns out to be
\begin{align}\label{N6}
\frac{-2\left[\theta+\ln\left(c_{4,3}
\left[\frac{C}{\varphi}\right]^\theta\right)\right]}
{\ln\left(c_{4,3}\left[\frac{C}{\varphi}\right]^\theta\right)} \geq 0.
\end{align}
In (\ref{N4}) we found that $\theta <0$, therefore, according to (\ref{er1-shartbt1}), one has
\begin{align}\label{afshin1}
\frac{1}{c_{4,3}}\left(\frac{\varphi}{C}\right)^\theta>1,
\end{align}
or equivalently
\begin{align}\label{afshin2}
0 < c_{4,3}\left(\frac{C}{\varphi}\right)^\theta <1
\end{align}
so
\begin{align}\label{afshin3}
\ln\left(c_{4,3}\left[\frac{C}{\varphi}\right]^\theta \right)<0.
\end{align}
Thus, (\ref{N6}) is not held here. Since the conditions (\ref{sh-cc-V-B}) and (\ref{N6}) which have been paired in the ``Class I'' in ref. \cite{hashem}, did not satisfy through the same conditions, hence we leave this set of conditions. Note that it means that if the reconstruction procedure is performed by decaying scalar field, then the potential (\ref{N5}) can never be `D.S.P.' type.\\

$\bullet$ \textbf{The second set of the potential conditions (Type-II of the scalar functions: Increasing scalar-field functions w.r.t. time):}

The condition $\mathfrak{B}[B,\gamma_{0}=0; V(B)] \geq 0$ for the potential (\ref{pot-cc-B}) is
\begin{align}\label{sh-cc-V-B-2}
\frac{\sqrt{\ln(B)}+\theta}{\theta \ln(B)} \geq 0.
\end{align}
Because it was obtained that $B>1$, hence for the positive amounts of $\theta$, this condition is true without any further assumption. For the negative values of $\theta$, this condition gives an upper bound for the scale factor, that is $B \leq \exp(\theta^2)$ and therefore $B \in (1,e^{\theta^2}]$. Another physical condition, which paired with this condition, is obtained by the use of $\mathfrak{B}[\varphi, \phi_{0}=0; V(\varphi)] \leq 0$ and (\ref{N5}) as
\begin{align}\label{N6-1}
\frac{-2\left[\theta+\ln\left(c_{4,3}
\left[\frac{C}{\varphi}\right]^\theta\right)\right]}
{\ln\left(c_{4,3}\left[\frac{C}{\varphi}\right]^\theta\right)} \leq 0.
\end{align}
According to (\ref{er1-shartbt1}), this condition is satisfied automatically for all real values of $\theta$ excluding zero. Therefore, both positive and negative amounts of $\theta$ are admissible, with the difference that in its negative values, the scale factor has an upper bound. Since $\theta$ is a constant, therefore if we accept the negative amounts of it, then indeed we tell that expanding of the universe shall end in the future. However, one may take $\theta \ggg$ to remove this restriction. It is worth noting that this bound coincides the theory that the geometry of the universe is, at least on a very large scale, elliptic. In a closed universe, gravity eventually stops the expansion of the universe, after which it starts to contract until all matter in the universe collapses to a point, a final singularity termed the ``Big Crunch'', the opposite of the ``Big Bang''.\\

For the potential (\ref{N5}), $\Gamma$ would be
\begin{align}\label{afshin5}
\Gamma=\frac{\theta^2+5 \theta \ln \left(c_{4,3}\left[\frac{C}{\varphi} \right]^{\theta} \right)+3 \ln^2 \left(c_{4,3}\left[\frac{C}{\varphi} \right]^{\theta} \right)}{2 \left(\theta+ \ln \left(c_{4,3}\left[\frac{C}{\varphi} \right]^{\theta} \right) \right)^2},
\end{align}
which may be written in the convenient form as
\begin{align}\label{afshin6}
\Gamma=\underbrace{\left(\frac{1}{2} \; \frac{\theta^2}{\ln^2(\varpi)}
+\frac{5}{2} \; \frac{\theta}{\ln (\varpi)}+\frac{3}{2} \right)}_{>\frac{3}{2}}
\underbrace{\left(\frac{\ln(\varpi)}{\theta + \ln(\varpi)} \right)^2}_{0< \text{between} <1},
\end{align}
where
\begin{align}\label{afshin7}
\varpi=c_{4,3}\left[\frac{C}{\varphi} \right]^{\theta}.
\end{align}
The behavior of the potential (\ref{N5}) is completely depended upon the amounts of $\theta$ and $\varpi$. In Table (\ref{tracker00}), it has been demonstrated that all three different behaviors of the potential (\ref{N5}) are possible for both positive and negative values of $\theta$. Note that pursuant to (\ref{er1-shartbt1}), for the positive and negative amounts of $\theta$, we have $\ln(\varpi)>0$ and $\ln(\varpi)<0$, respectively. Therefore, according to the Table (\ref{tracker00}) and due to the increasing nature of $\ln(\varpi)$ with time, if we set $|\theta|< 1.618 \; |\ln(\varpi)|$ as an initial condition for both signs of $\theta$, then three different dynamics of potential namely `Thawing', `Scaling', and `Tracker' are achieved, respectively, as the universe ages.\\

\begin{table*}
\caption{Behaviors of the potential $V(\varphi)=\tilde{V}_{0cc}\varphi^{-2} \ln^{-2}(\varpi)$} \label{tracker00}
\centering
\begin{tabular}{l l c c}
\toprule[1.7pt]
Sign of $\theta$ \quad & Sign of $\ln(\varpi)$ & Condition & Cosmological dynamics of the potential
\\ [0.5ex]
\toprule[1.4pt] 
$\theta>0$ & $\ln(\varpi)>0$ & $\theta > \left(\frac{1}{2}+\frac{\sqrt{5}}{2} \right)\ln(\varpi)$ & $\Gamma <1$; \; Thawing\\ [0.5ex]
\toprule[0.7pt]
$\theta<0$ & $\ln(\varpi)<0$ & $\theta < \left(\frac{1}{2}+\frac{\sqrt{5}}{2} \right)\ln(\varpi)$ & $\Gamma <1$; \; Thawing\\ [0.5ex]
\toprule[0.7pt]
$\theta>0$ & $\ln(\varpi)>0$ & $\theta= \left(\frac{1}{2}+\frac{\sqrt{5}}{2}\right)\ln(\varpi)$ & $\Gamma =1$; \; Scaling\\ [0.5ex]
\toprule[0.7pt]
$\theta <0$ & $\ln(\varpi) <0$ & $\theta= \left(\frac{1}{2}+\frac{\sqrt{5}}{2}\right)\ln(\varpi)$ & $\Gamma =1$; \; Scaling\\ [0.5ex]
\toprule[0.7pt]
$\theta > 0$ & $\ln(\varpi) >0$ & $\theta< \left(\frac{1}{2}+\frac{\sqrt{5}}{2} \right)\ln(\varpi) $ & $\Gamma >1$; \; Tracker\\ [0.5ex]
\toprule[0.7pt]
$\theta < 0$ & $\ln(\varpi) <0$ & $\theta> \left(\frac{1}{2}+\frac{\sqrt{5}}{2} \right)\ln(\varpi) $ & $\Gamma >1$; \; Tracker\\ [0.5ex]
\toprule[1.7pt]
\end{tabular}
\end{table*}

For the cosmological constant case, without knowing the form of the scale factor and also scalar field in terms of the time, we found out that the positive and negative amounts of $\theta$ form two separate families of admissible solutions. So, one may examine some of the scale factor models to see they belong to which of the two. On the other hand, since the `cosmological constant' case has been suggested for the accelerating era, hence the scale factors which presented for this era are only testable.\\
It is interesting that we can understand one of the admissible ranges of $\theta$ via another way:\\
The known scale factor for the dark energy dominated
era is $e^{H_{0}t}$, where the coefficient $H_{0}$ in the exponential is the Hubble constant. We may anticipate that this form of the scale factor and the one obtained, (\ref{moa-c-1}), must be proportional:
\begin{align}\label{yek.1}
& a_{ave.}= B^{\frac{(m+2)}{3}} \propto \exp\left[H_{0}t \right] \nonumber \\ & \Longrightarrow \left(\exp \left[\left(\ln \left(\frac{1}{c_{4,3}}\left[\frac{\varphi}{C} \right]^{\theta} \right)\right)^2 \right]\right)^{\frac{(m+2)}{3}} \propto \exp\left[H_{0}t \right]
\end{align}
Therefore, the scalar field is achieved as
\begin{align}\label{yek.2}
\varphi(t) \propto C\left(c_{4,3}\exp\left[\sqrt{\frac{3H_{0} \; t}{m+2}} \right] \right)^{\frac{1}{\theta}}.
\end{align}
As mentioned earlier, in this paper, we restrict ourselves to a decaying scalar field versus time, i.e. $\mathfrak{B}[t,0;\varphi(t)]<0$. Here, it gives $\theta<0$ which agrees with the thing was earned, and as discussed above it implies the `tracker' nature for the potential (\ref{N5}). In view of (\ref{er1-shartbt1}) and (\ref{e111}), negative amounts of $\theta$ put an upper and lower bounds on the scalar field as $\varphi <C c^{1/\theta}_{4,3}$ and $\varphi > C c^{1/\theta}_{4,3}$ when $\varphi>0$ (therefore $C>0$) and $\varphi<0$ (therefore $C<0$), respectively, which regarding (\ref{yek.2}), both yield a trivial result: $\sqrt{3H_{0} \;t /(m+2)}>0$. One of the confirmations of $\mathfrak{B}[t,0;\varphi(t)]<0$ is that if unlike its result namely $\theta<0$, one goes with the positive amounts of $\theta$ and the scale factor (\ref{yek.2}) ahead, then encounters with a strange outcome: $\sqrt{3H_{0} \;t /(m+2)}<0$. As we observe, the physical ranges of all constant parameters are tightly coupled together.\\

\noindent\rgoldmybox{red}{$\textcolor[rgb]{0.00,0.25,0.50}{\mathbf{\ast} \textbf{2- The Second Way $l_{1} \neq l_{2}$:}}$}\\

Multiplying both sides of eq. (\ref{v28.1}) by $M(B)$, one has
\begin{align}\label{v28.2}
l_{1}\left(M_{,BB}+\frac{M_{,B}}{2B\ln(B)}\right)
+l_{2}\frac{M_{,B}}{B}+l_{3}\frac{M}{B^2\ln(B)}=0.
\end{align}
Assuming $l_{1} \neq l_{2}$, this equation generates the following function:
\begin{align}\label{v29}
M(B) &=c_{8}\sqrt{\ln(B)} \nonumber \\ & \times \mathbf{Kummer}\mathbb{M}
\left(\frac{l_{1}-l_{2}-2l_{3}}{2l_{1}-2l_{2}},\frac{3}{2}
,\frac{\left(l_{1}-l_{2}\right)\ln(B)}{l_{1}}\right) \nonumber \\
&+c_{9}\sqrt{\ln(B)} \nonumber \\ & \times
\mathbf{Kummer}\mathbb{U}\left(\frac{l_{1}-l_{2}-2l_{3}}{2l_{1}-2l_{2}},\frac{3}{2}
,\frac{\left(l_{1}-l_{2}\right)\ln(B)}{l_{1}}\right),
\end{align}
where $c_{8}$ and $c_{9}$ are integration constants, and $\mathbf{Kummer}\mathbb{M}(a,b,z)$ and $\mathbf{Kummer}\mathbb{U}(a,b,z)$ are Kummer functions. Note that the arguments of both Kummer functions are the same. Obviously, $M(B)$ must be nonzero. Therefore, if the conditions $$0\leq \frac{3}{2} - \frac{l_{1}-l_{2}-2l_{3}}{2l_{1}-2l_{2}}\leq 1$$ or equivalently $$-1 \leq \frac{l_{3}}{l_{1}-l_{2}} \leq 0,$$ and $B \neq 0, 1$ are imposed, then the solution (\ref{v29}) will be nonzero and also nonsingular in the throughout of evolution range of $B$ (See `Zeros of $\mathbf{Kummer}\mathbb{M}(a,b,z)$' and `Zeros of $\mathbf{Kummer}\mathbb{U}(a,b,z)$' in ref. \cite{book1}). Note that at the points $B=0, 1$ there is no singularity problem for $\mathbf{Kummer}\mathbb{M}(a,b,z)$, rather, due to $\sqrt{\ln(B)}$, we have put $B\neq 0, 1$. Furthermore, again we set $B>1$ for the foregoing reasons (i.e. having real-valued Hubble parameter and etcetera). Manifestly, this solution is not invertible. However, one can do all steps numerically. But, we would like to perform it through an analytical way. So, let us carry two reasonable choices out for simplification
\begin{align}\label{aaa1}
\frac{l_{1}-l_{2}-2l_{3}}{2l_{1}-2l_{2}}=\frac{3}{2}, \qquad c_{9}=0.
\end{align}
Therefore, the solution (\ref{v29}) turns out to be
\begin{align}\label{aaa2}
M(B) & =c_{8}\sqrt{\ln(B)}\; \; \mathbf{Kummer}\mathbb{M}
\left(\frac{3}{2},\frac{3}{2} ,\frac{\left(l_{1}-l_{2}\right)\ln(B)}{l_{1}}\right)
\nonumber \\ & =c_{8}\sqrt{\ln(B)}\; \; B^{\frac{l_{1}-l_{2}}{l_{1}}}
\end{align}
where we have used the connection formula \cite{book1,book2}
\begin{align}\label{K19}
\mathbf{Kummer}\mathbb{M}(a,a,z)=e^z.
\end{align}
Regarding (\ref{shekle M}), the corresponding scalar field would be
\begin{align}\label{aaa3}
\varphi(B)=C \left(c_{8}\sqrt{\ln(B)}\; \; B^{\frac{l_{1}-l_{2}}{l_{1}}} \right)^{\frac{1}{\theta}}.
\end{align}
It may be written as
\begin{align}\label{aaa4}
\ln(B)\; \; B^{\frac{2\left(l_{1}-l_{2} \right)}{l_{1}}}=\frac{1}{c^2_{8}}\left(\frac{\varphi}{C}\right)^{2 \theta}.
\end{align}
Now, we can invert eq. (\ref{aaa3}) by solving (\ref{aaa4}). Limpidly, this equation has a strong connection with the well-known Lambert equation. Solving it gives the form of the scale factor in terms of the scalar field:
\begin{align}\label{aaa5}
B(\varphi)=\exp & \left[\frac{l_{1}}{2\left(l_{2}-l_{1}\right)} \; 
\mathbf{LambertW}
\left(\frac{2\left(l_{2}-l_{1}\right)}{l_{1}c^2_{8}}\left[\frac{\varphi}{C} \right]^{2\theta} \right) \right].
\end{align}
Using (\ref{res sol for NS2}), (\ref{v26-1}), and (\ref{aaa3}) in (\ref{v8}), one obtains
\begin{align}\label{aaa6}
V(B)=V_{0cc2} \left(B^{\frac{2\left(l_{1}-l_{2}\right)}{l_{1}}} \ln(B) \right)^{\frac{1}{\theta}} \left(\zeta_{1}\ln(B)+\zeta_{2}+\frac{\zeta_{3}}{\ln(B)} \right),
\end{align}
in which
\begin{align}\label{aaa7}
V_{0cc2}=& \frac{C^2 H^2_{0c} c^{\frac{2}{\theta}}_{8}}{8 \theta^2 l^2_{1}}\left[\frac{2 \rho_{0}}{m+3} \left(2-\frac{m^2+m+4}{2m+1}\right) \right],\nonumber \\
\zeta_{1}=&4\left[4(2m+1)f_{0} \theta^2+4(2mg_{0}+1)h_{0}\theta-\omega_{0} \right]l^2_{1} \nonumber \\ &-8\left[2(2mg_{0}+1)h_{0}\theta-\omega_{0} \right]l_{1}l_{2}-4\omega_{0}l^2_{2},\nonumber \\
\zeta_{2}=&4 \left[2(2mg_{0}+1)h_{0}\theta-\omega_{0} \right]l^2_{1}+4l_{1}l_{2}\omega_{0},\nonumber \\
\zeta_{3}=& -\omega_{0} l^2_{1}.
\end{align}
Eqs. (\ref{aaa5}) and (\ref{aaa6}) give the potential in terms of the scalar field:
\begin{align}\label{aaa8}
V(\varphi)=\frac{-V_{0cc3}}{\varphi^2}\left[\zeta_{4}
\mathbf{W}^{\frac{2}{\theta}+1}+\zeta_{5}\mathbf{W}
^{\frac{2}{\theta}}+\zeta_{6}\mathbf{W}^{1-\frac{2}{\theta}}\right],
\end{align}
where
\begin{align}\label{aaa9}
&\mathbf{W}=\mathbf{LambertW}
\left(\frac{2(l_{2}-l_{1})}{l_{1}c^2_{8}}
\left[\frac{\varphi}{C}\right]^{2\theta} \right),\nonumber \\
&V_{0cc3}=\frac{1}{2}V_{0cc2}C^2 \left(\frac{l_{1} c_{8}}{2(l_{1}-l_{2})}\right)^{\frac{2}{\theta}}, \nonumber \\
&\zeta_{4}=\frac{l^{\theta}_{1}}{l_{1}-l_{2}}\zeta_{1}, \quad
\zeta_{5}=-6^{\frac{-1}{\theta}}\zeta_{2},\quad
\zeta_{6}=4\left(1-l^{1-\frac{4}{\theta}}_{1} \right)\zeta_{3}.
\end{align}
The Taylor series of Lambert function around zero can be found using the Lagrange inversion theorem and is given by
\begin{align}\label{aaa10}
\mathbf{LambertW}(x)=\sum_{j=1}^{\infty}\frac{(-j)^{j-1}}{j!}x^j.
\end{align}
So, the potential (\ref{aaa8}) may be written as
\begin{align}\label{aaa11}
V(\varphi)=\frac{V_{0cc3}}{\varphi^2} \sum_{j=1}^{\infty}k_{j}\left( \zeta_{4}\varphi^{2\left(2+\theta \right)j}+\zeta_{5}\varphi^{4j}+\zeta_{6}\varphi^{2\left(\theta -2 \right)j}\right),
\end{align}
in which
\begin{align}\label{aaa12}
k_{j}=\frac{j^{j-1}}{j!} \left(\frac{2 (l_{1}-l_{2})}{l_{1}C^{2\theta}c^2_{8}}\right)^j.
\end{align}
It seems that the form (\ref{aaa11}) is better and more prevalent than (\ref{aaa8}). Respecting to the range of $\theta$, one of the terms $\zeta_{4}\varphi^{2\left(2+\theta \right)j}$ and $\zeta_{6}\varphi^{2\left(\theta -2 \right)j}$ in the series is removable.\\
Here, we are again able to discuss the physically admissible domains of parameters. According to $\mathfrak{B}[\varphi, \beta_{0}=0; B(\varphi)]<0$, the scale factor presented in (\ref{aaa5}) must satisfy this condition:
\begin{align}\label{aaa13}
\left(\frac{l_{1}\theta}{l_{2}-l_{1}}\right)
\left(\frac{\mathbf{W}}{1+\mathbf{W}} \right)<0.
\end{align}
\begin{figure*}
\centering
\includegraphics[width=3 in, height=2 in]{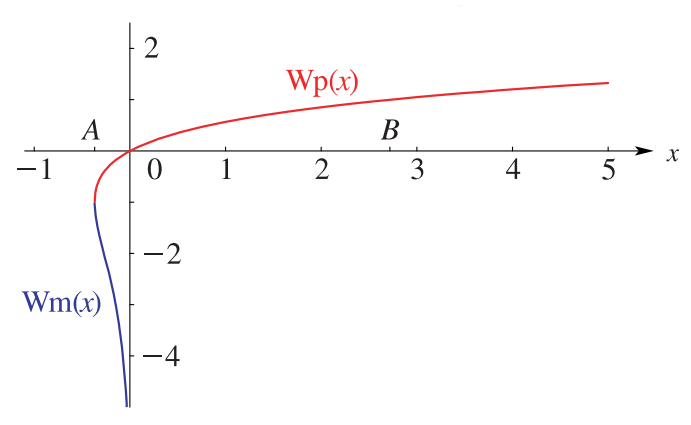}\\
\caption{Branches $Wp(x)$ (the positive branch) and $Wm(x)$ (the negative branch) of the Lambert W-function. $A$ and $B$ denote the points $-1/e$ and $e$, respectively, on the $x$-axis \cite{book1}.}\label{fig1}
\end{figure*}
First of all, see the plot of $\mathbf{LambertW}$ function (Fig. (\ref{fig1})). As we observe, the positive branch is more physical than one another, especially for its widely finite range. So, its argument should here be held positive, viz.,
\begin{equation}\label{aaa14}\begin{split}
\frac{2(l_{2}-l_{1})}{l_{1}c^2_{8}}\left(\frac{\varphi}{C} \right)^{2\theta}>0 \; \Longrightarrow  \left\{
                                                  \begin{array}{ll}
                                                    1:\;  \frac{l_{2}-l_{1}}{l_{1}}>0;\\ \\
                                                    2:\; \frac{\varphi}{|\varphi|}
\frac{C}{|C|}>0 .
                                                  \end{array}
                                                \right.
\end{split}\end{equation}
Since the sign of the scalar field is indeterminate, and on the other hand, we do know the amount of $\theta$ --- only for its integer amounts, $2\theta$ is even, not in general ---, so it would be better that the sign of $C$ be fixed by $\varphi$; the second resultant condition in (\ref{aaa14}) has come from here. From (\ref{aaa13}) and (\ref{aaa14}), it appears that $\theta < 0$. The first outcome condition in (\ref{aaa14}) implies that $(l_{2}/l_{1})>1$, hence $l_{2}$ and $l_{1}$ have the same sign and therefore $\left|l_{2} \right| > \left|l_{1} \right|$.\\
In view of $\mathfrak{B}[B,\alpha_{0}=0; \varphi(B)]<0$ and (\ref{aaa3}), we gain another condition for the ratio $l_{2}/l_{1}$ as
\begin{align}\label{aaa15}
\frac{l_{2}}{l_{1}} <\frac{1}{2\ln(B)}+1.
\end{align}
Combining this condition with the previous bound $(l_{2}/l_{1})>1$, we get
\begin{align}\label{aaa16}
1<\frac{l_{2}}{l_{1}} <\frac{1}{2\ln(B)}+1.
\end{align}
The conditions (\ref{chaharshart1})-(\ref{chaharshart2}) also confirm these conditions and do not add further conditions and bounds.\\
The potential conditions do not present straightforward constraints for the parameters, because of the form of potential. The conditions are long; see
\begin{align}\label{aaa17}
& \mathfrak{B}[B,0; V(B)] \nonumber \\ &=\bigg\{\theta l_{1}\ln(B) \left[\zeta_{1}\ln^2(B)+\zeta_{2}\ln(B)+\zeta_{3}\right] \bigg\}^{-1} \nonumber \\ & \times
\biggl\{2\zeta_{1}(l_{1}-l_{2})\ln^3(B)+\left[\left((1+\theta)\zeta_{1}
+2\zeta_{2}\right)l_{1} \right. \nonumber \\ & \left. -2l_{2}\zeta_{2} \right]\ln^2(B)+\left[(\zeta_{2}+2\zeta_{3})l_{1}-2\zeta_{3}l_{2} \right]\ln(B) \nonumber \\ & -\zeta_{3}l_{1}(\theta-1) \biggl\},
\end{align}
and
\begin{align}\label{aaa18}
&\mathfrak{B}[\varphi,0; V(\varphi)]
\nonumber \\ & =\bigg\{ \left(1+\mathbf{W}\right)\left(\zeta_{4}\mathbf{W}^
{\frac{4}{\theta}+1}+\zeta_{5}
\mathbf{W}^{\frac{4}{\theta}}+\zeta_{6}\mathbf{W}\right) \bigg\}^{-1} \nonumber \\
&\times \biggl\{-2\zeta_{4}\mathbf{W}^{\frac{4}{\theta}+2}
+\left[(2\theta+2)\zeta_{4}-2\zeta_{5} \right]\mathbf{W}^{\frac{4}{\theta}+1}
\nonumber \\ & +2\zeta_{5}\mathbf{W}^{\frac{4}{\theta}}
+2\zeta_{6}\left(\theta-3-\mathbf{W} \right)\mathbf{W} \biggl\}.
\end{align}
Only in plotting or in the data analysis, these would be helpful. Because we did not set a special value for each of the constant parameters, hence we ignore to proceed further and for this problem, we have not written the $\geq$ and $\leq$ symbols in the above conditions. As mentioned in ref. \cite{hashem}, the conditions have been paired, so we must take care when one condition is nonpositive, another condition must be nonnegative, and vice-versa.\\
Considering the function $\Gamma$ is also ineffectual struggle when we do not determine exactly the amounts of constants.\\

The taken procedure at the end of the first way, may here be performed for (\ref{aaa5}) as well. Therefore, for the reason aforementioned, we start with
\begin{align}\label{Ali1}
\left(\exp\left[\frac{l_{1}}{2\left(l_{2}-l_{1}\right)}\mathbf{LambertW}
\left(\frac{2\left(l_{2}-l_{1}\right)}{l_{1}c^2_{8}}\left[\frac{\varphi}{C} \right]^{2\theta} \right) \right]\right)^{\frac{m+2}{3}} =\exp \left[H_{0}t \right].
\end{align}
Because finding the physically admissible domains out for the parameters is our objective, thus (\ref{Ali1}) must hold for all its true regions, hence we exploite the Taylor expansion for $\mathbf{LambertW}$ function up to the first term yielding
\begin{align}\label{Ali2}
\frac{1}{c^2_{8}}\left[\frac{\varphi}{C} \right]^{2\theta}=\frac{3H_{0}t}{m+2}.
\end{align}
Obviously, since we do not know that $\theta$ is an integer or not, so for the sake of precaution, the condition which was found earlier (See the second condition in (\ref{aaa14})), viz.,
\begin{align}\label{Ali3}
\frac{\varphi}{\left| \varphi \right|}\frac{C}{\left| C \right|}>0,
\end{align}
should be adopted. Also, the scalar field reads
\begin{align}\label{Ali4}
\varphi(t)=C\left(\frac{3 c^2_{8} H_{0}}{m+2}\right)^{\frac{1}{2\theta}}t^{\frac{1}{2\theta}},
\end{align}
which according to $\mathfrak{B}[t,0;\varphi(t)]<0$, leads to
\begin{align}\label{Ali5}
\theta<0,
\end{align}
as it emerged previously (See some lines after (\ref{aaa14})).\\

\noindent\qrmybox{red}{\paragraph{\textbf{The second case: FRW ($A=B$; $m=1$):}}}\\

In the FRW case, i.e. $m=1$, by inserting $P=-\rho$ into the right side of eq. (\ref{hard1}) and combining with (\ref{hard2}), we arrive at $\dot{H}=0$ ($\mathfrak{B}[t,0;H(t)]=0$) or equivalently $H_{,B}=0$ ($\mathfrak{B}[B,0;H(B)]=0$) because of $\dot{H}=BHH_{,B}$. Therefore, according to ``Table (III)'' or ``Figure 1'' in ref. \cite{hashem}, and $\mathfrak{B}[t,0;H(t)]$ or $\mathfrak{B}[B,0;H(B)]$ we learn that this case gives `Accelerated expansion' and `Phase transition point (Phantom divide line)'. It is worth mentioning that in this case, the Hubble parameter is constant, so the deceleration parameter turns out to be minus one which coincides with its present amount according to the observational data.\\
For this case, the basic equation (\ref{v15}) becomes
\begin{align}\label{aaa19}
c_{10}\frac{M_{,BB}}{M}+c_{11}\frac{1}{B}\frac{M_{,B}}{M}=0,
\end{align}
in which
\begin{align}\label{aaa20}
c_{10}&=\frac{2(2g_{0}+1)}{3\omega_{0}+4h_{0}(2g_{0}+1)},\nonumber \\
c_{11}&=\left(\frac{6}{3\omega_{0}+4h_{0}(2g_{0}+1)}\right)
\left(4\frac{f_{0}}{h_{0}}-3g_{0}\right).
\end{align}
It leads to the following solution
\begin{align}\label{aaa21}
M(B)=c_{12}+c_{13}B^{\frac{c_{10}-c_{11}}{c_{10}}},
\end{align}
where $c_{12}$ and $c_{13}$ are the integration constants. For simplicity, let us take $c_{12}=0$. From eq. (\ref{shekle M}) one has
\begin{align}\label{aaa22}
\varphi(B)=C \left(c_{13} B^{\frac{c_{10}-c_{11}}{c_{10}}}\right)^{\frac{1}{\theta}},
\end{align}
and consequently
\begin{align}\label{aaa23}
B(\varphi)=\left[\frac{1}{c_{13}}\left(\frac{\varphi}{C} \right)^{\theta} \right]^{\frac{c_{10}}{c_{10}-c_{11}}}.
\end{align}
For these forms of the $B(\varphi)$ and $\varphi(B)$, the conditions $\mathfrak{B}[\varphi, \beta_{0}=0; B(\varphi)]<0$ and $\mathfrak{B}[B, \alpha_{0}=0; \varphi(B)]<0$ imply
\begin{align}\label{aaa24}
\frac{\theta c_{10}}{c_{10}-c_{11}}<0,
\end{align}
and
\begin{align}\label{aaa25}
\underbrace{\left(\frac{c_{10}-c_{11}}{\theta c_{10}} \right)}_{<0; \text{ cf.(\ref{aaa24})}} \left(\frac{c_{13}}{c_{12}} \right)<0 \quad \Longrightarrow \quad \frac{c_{13}}{c_{12}}>0,
\end{align}
respectively. Therefore, the latter one condition reveals $|c_{13}|>|c_{12}|$. The conditions (\ref{chaharshart1})-(\ref{chaharshart2}) also give (\ref{aaa24}). With (\ref{aaa22}) at hand, the potential in terms of the scale factor is given by (\ref{v8}) as
\begin{align}\label{aaa26}
V(B)=V_{0cc4}C^2 c^{\frac{2}{\theta}} B^{\frac{2(c_{10}-c_{11})}{\theta c_{10}}},
\end{align}
where
\begin{align}\label{aaa27}
V_{0cc4}=& \frac{6H^2_{0cc}}{\theta^2 c^2_{10}}\biggl[\left(\theta^2 f^2_{0}+\frac{2}{3}h_{0}\theta\left(g_{0}+\frac{1}{2}\right)-\frac{1}{12}
\omega_{0} \right)c^2_{10} \nonumber \\ &-\frac{2}{3}c_{10}c_{11} \left(h_{0}\theta\left( g_{0}+\frac{1}{2}\right)-\frac{1}{4}\omega_{0}\right)-\frac{1}{12}c^2_{11}
\omega_{0} \biggl],
\end{align}
in which $H_{0cc}$ is the Hubble constant. Correspondingly, from (\ref{aaa23}) and (\ref{aaa26}) we gain the quadratic form for the scalar potential:
\begin{align}\label{aaa28}
V(\varphi)=V_{0cc4}\; \; \varphi^2.
\end{align}
The significance and beauty of this result lie with the fact that it has been emerged by the use of Noether symmetry approach with worthwhile results in ref. \cite{beh-intervention}.\\
Because we have
\begin{align}\label{aaa29}
\mathfrak{B}[\varphi, \phi_{0}=0; V(\varphi)]=2 \geq 0,
\end{align}
thus its paired condition namely $\mathfrak{B}[B, \gamma_{0}=0; V(B)] \leq 0$ must be taken into account:
\begin{align}\label{aaa30}
\frac{c_{10}-c_{11}}{\theta c_{10}} \leq 0
\end{align}
which is compatible with (\ref{aaa24}), excluding that zero is not admissible. Note that the common domain must be established (i.e. $\frac{c_{10}-c_{11}}{\theta c_{10}} < 0$ is true, not $\frac{c_{10}-c_{11}}{\theta c_{10}} \leq 0$). Limpidly, also one has $\theta \neq 0$, $c_{10} \neq 0$, $c_{12} \neq 0$, $c_{13} \neq 0$, and $c_{10} \neq c_{11}$. Under the reconstruction with a decaying scalar field, the potential (\ref{aaa28}) would always be of `D.S.P.'-type because of (\ref{aaa29}).\\
Clearly, the potential (\ref{aaa28}) is of `Thawing'-type, because of $\Gamma=1/2$.\\

\noindent\Rrmybox{}{\subsubsection{\textbf{Modified Chaplygin gas}}}\vspace{5mm}

In the simplest case, (standard) Chaplygin gas is a perfect fluid characterized by the equation of state
\begin{align}\label{ccc1}
P=-\frac{C_{chg}}{\rho},
\end{align}
where $C_{chg}$ is a positive constant. Because the Chaplygin gas formally carries negative pressure, it is exploited in cosmology to describe a transition from a decelerated cosmological expansion to the present epoch of a cosmic acceleration. Also, it describes a unification of dark matter and dark energy.

Another model that has been discussed in some details in ref. \cite{GCG} is the generalized Chaplygin gas that has two free parameters:
\begin{align}\label{ccc2}
P=-\frac{C_{gchg}}{\rho^{\alpha_{1}}},
\end{align}
where $0<\alpha_{1}\leq 1$ and $C_{gchg}$ is a positive constant. Clearly, when $\alpha_{1}=1$, the generalized Chaplygin gas (\ref{ccc2}) reduces to the Chaplygin gas (\ref{ccc1}).

Within the framework of Friedman-Robertson-Walker (FRW) cosmology, a model called modified Chaplygin gas has been proposed by H.B. Benaoum \cite{Benaoum}, in which his principal assumption is that the energy density $\rho$ and pressure $P$ are related by the following equation of state:
\begin{align}\label{ccc3}
P=\sigma_{1}\rho - \frac{\sigma_{2}}{\rho^{\nu}}.
\end{align}
In ref. \cite{Benaoum}, he adopted three assumptions: $\sigma_{1}$ and $\sigma_{2}$ are positive constants and $\nu \geq 1$. But, these assumptions seem to be incorrect, because of the ``correspondence principle''. According to the ``correspondence principle'', a new scientific theory (or a generalized theory) must reduce to an earlier scientific theory in appropriate circumstances. This requires that the new theory must explain all the phenomena under circumstances for which the preceding theory was known to be valid, the ``correspondence limit''. Here, indeed (\ref{ccc3}) is a mixture of two things: perfect fluid (\ref{vgb1}) and generalized Chaplygin gas (\ref{ccc2}). Therefore, the assumptions should at least be taken as: $0\leq \sigma_{1}\leq 1$, $0< \nu\leq 1$, and $\sigma_{2}$ is a nonnegative constant (i.e. $\sigma_{2}\geq 0$). According to the Table (\ref{bestfitted}), it is revealed that $\sigma_{1,min.}=-0.018$, $\sigma_{1,max.}=+0.207$, $\nu_{1,min.}=-0.1832$, and $\nu_{1,max.}=+1.724$. Therefore, it would be good enough to modify the domains of $\sigma_{1}$ and $\nu$ again as $-0.02< \sigma_{1}<0.21$ and $-0.2< \nu < 1.8$. However, the range of $\nu$ may be restricted further, because, in ref. \cite{darksector} it has been demonstrated that the generalized Chaplygin gas correctly describes the cosmological dark sector when $|\nu|\lesssim 0.05$, i.e. it has to be very close to the $\Lambda\text{CDM}$ model which corresponds to $\nu=0$. Anyway, with these new conditions, some well-known cases would be taken into account:
\begin{enumerate}
  \item When $\sigma_{1}=\sigma_{2}=0$ $\Longrightarrow$ $P=0$ (Matter-dominated era: Pressure-less matter: Dust; Note that here $\rho \neq 0$).
  \item When $\sigma_{1}=\frac{1}{3}$ and $\sigma_{2}=0$ $\Longrightarrow$ $P=\frac{1}{3}\rho$ (Radiation-dominated era).
  \item When $0 < \sigma_{1} < 1$ and $\sigma_{2}=0$ $\Longrightarrow$ $P=\sigma_{1} \rho$ (Bartropic fluid).
  \item When $\sigma_{1}=0$ and $\sigma_{2} \neq 0$ $\Longrightarrow $ $P=-\frac{\sigma_{2}}{\rho^{\nu}}$ (Generalized Chaplygin gas).
  \item When $\sigma_{1}=0$, $\sigma_{2} \neq 0$, and $\nu=1$ $\Longrightarrow$ $P=-\frac{\sigma_{2}}{\rho}$ (Standard Chaplygin gas).
  \item When $\sigma_{1}=0$, and $\nu=0$ $\Longrightarrow$ $P=-\sigma_{2}$ ($\Lambda\text{CDM}$ model)
\end{enumerate}

\begin{table*}
\caption{Constraints on modified Chaplygin gas} \label{bestfitted}
\centering
\begin{tabular}{c c c}
\toprule[1.7pt]
$\sigma_{1}$ &$\nu$ & The constraints arise from...
\\ [0.5ex]
\toprule[1.4pt]
$0.085$ & $1.724$ & The best fitted parameters \cite{ppllbb1} \\ [0.5ex]
\toprule[0.7pt]
$ 0.061 \pm 0.079$ & $0.053 \pm 0.089$ & Constitution + CMB + BAO \cite{ppllbb2}\\ [0.5ex]
\toprule[0.7pt]
 $0.110 \pm 0.097$ & $0.089 \pm 0.099$ & Union + CMB + BAO \cite{ppllbb2}\\ [0.5ex]
\toprule[0.7pt]
 $0.00189^{+0.00583}_{-0.00756}$ & $0.1079^{+0.3397}_{-0.2539}$ & Markov Chain Monte Carlo method at $1\sigma$ level \cite{ppllbb3}\\ [0.5ex]
\toprule[0.7pt]
  $0.00189^{+0.00660}_{-0.00915}$ & $0.1079^{+0.4678}_{-0.2911}$ & Markov Chain Monte Carlo method at $1\sigma$ level \cite{ppllbb3}\\ [0.5ex]
\toprule[1.7pt]
\end{tabular}
\end{table*}
Substituting $P$ given in (\ref{ccc3}) into the conservation equation (\ref{chegali}), we get the evolution of the energy density of the modified Chaplygin gas in terms of the scale factor as
\begin{align}\label{ccc4}
\rho=\left(\frac{\sigma_{2}}{1+\sigma_{1}}
+\frac{b_{0}}{B^{(1+\sigma_{1})(1+\nu)(m+2)}} \right)^{\frac{1}{1+\nu}},
\end{align}
where $b_{0}$ is an integration constant which may be taken as the matter content of the universe at present.\\

Replacing (\ref{ccc3}) on the right-hand side of (\ref{hard1}) and combining with (\ref{hard2}) we obtain
\begin{align}\label{ccc5}
k_{1} H^{2+2\nu}_{2}+k_{2}BH^{1+2\nu}_{2}H_{2,B}-k_{3}=0,
\end{align}
where
\begin{align}\label{ccc6}
k_{1}&=\left(\frac{m^2+m+4}{2} \right)+\sigma_{1}(1+2m), \nonumber \\
k_{2}&=\frac{m+3}{2}, \nonumber \\
k_{3}&=\frac{\sigma_{2}}{(1+2m)^{\nu}}.
\end{align}
In (\ref{ccc5}), we have used the relation $\dot{H}_{2}=BH_{2}H_{2,B}$. Solving eq. (\ref{ccc5}) gives the form of the Hubble parameter in terms of the scale factor:
\begin{align}\label{ccc7}
H_{2}=\left(c_{14}B^{\frac{-2k_{1}}{k_{2}}(1+\nu)}+\frac{k_{3}}{k_{1}} \right)^{\frac{1}{2(1+\nu)}},
\end{align}
where $c_{14}$ is an integration constant. Setting $c_{14}=1/k^2_{1}$, one consequently has
\begin{align}\label{ccc8}
\frac{H_{2,B}}{H_{2}}=\frac{-1}{k_{4}B+k_{5}B^{k_{6}+1}},
\end{align}
in which
\begin{align}\label{ccc9}
k_{4}&=\frac{k_{2}}{k_{1}}, \nonumber \\
k_{5}&=k_{2}k_{3}, \nonumber \\
k_{6}&=2(1+\nu)\frac{k_{1}}{k_{2}}=\frac{2(1+\nu)}{k_{4}}.
\end{align}
Note that $k_{i}>0$ when $i=1,2,4,6$ and $k_{i}\geq 0$ for $i=3,5$. In what follows, in order to achieve the explicit forms of functions and amounts, we must multiply or divide by $k_{3}$ and $k_{5}$, hence from now on, we impose $k_{3}>0$ and $k_{5}>0$. The expanding nature of the universe for the amounts of the scale factor in each time of the era of interest implies
\begin{align}\label{ex}
B>(-k_{1}k_{3})^{\frac{-k_{2}}{2k_{1}(1+\nu)}}
=(-k_{1}k_{3})^{\frac{-1}{k_{6}}},
\end{align}
where we have used $\mathfrak{B}[t,0;B(t)]>0$ and (\ref{ccc7}).\\
Utilizing (\ref{ccc8}), the function $\mathfrak{B}[a,0;H(a)]$ turns out to be
\begin{align}\label{109047}
\mathfrak{B}[a,0;H(a)]=\frac{-3}{(m+2)} \frac{1}{k_{4}+k_{5}B^{k_{6}}}.
\end{align}
According to ``Table (III)'' or ``Figure 1'' in ref. \cite{hashem}, eq. (\ref{109047}) identifies the amounts of the scale factor for the different states of the universe. Setting
\begin{equation}\begin{split}\label{tttable}
\vartheta_{1}&=\left[\frac{1}{k_{5}}\left(\frac{3}{(m+2)}-k_{4} \right)\right]^{\frac{1}{k_{6}}},\\
\vartheta_{2}&=\left (-k_{1}k_{3}\right)^{\frac{-k_{2}}{2k_{1}(1+\nu )}},
\end{split}\end{equation}
all possible behaviors have been presented in Table (\ref{table1-dec1}). As we observe in Table (\ref{table1-dec1}), phase transition from a quintessence phase to a phantom phase cannot occur at all, because, from physical point of view, the function $\mathfrak{B}[a,0;H(a)]$ cannot be zero. If one sets $B>0$, $k_{6}=\text{odd}$, and $B<\vartheta_{1}$ at first, then the universe poses from a decelerated expansion to an accelerated expansion as it ages and within this process, the phase of the universe remains in quintessence phase. An accelerated expansion which be in phantom phase exists when one considers the negative amounts of the scale factor $B$ together with the conditions $k_{6}=\text{odd}$ \; \& \; $k_{4}k_{6}=\text{even}$ \; \& \; $k_{4}+k_{5}B^{k_{6}}<0$ \; \& \; $\vartheta_{2}<B<\vartheta_{1}$.
\begin{table*}
\caption{Different behaviors of expansion of the universe and their corresponding amounts of the scale factors in modified Chaplygin gas case} \label{table1-dec1}
\centering
\begin{tabular}{c c}
\toprule[1.7pt]
\textbf{Conditions}&\textbf{The status of the universe}
\\ [0.5ex]
\toprule[1.4pt]
$k_{6}=\text{odd}$ \; \& \;$B>0$\; \& \; $B>\vartheta_{1}$ & \tiny{Accelerated expansion and Quintessence phase}\\ [1.3ex]
   \toprule[0.7pt]
$k_{6}=\text{odd}$ \; \& \;$B>0$ \; \& \; $B=\vartheta_{1}$ & \tiny{The inflection point and Quintessence phase}\\ [1.3ex]
    \toprule[0.7pt]
$k_{6}=\text{odd}$ \; \& \;$B>0$ \; \& \; $B<\vartheta_{1}$ & \tiny{Decelerated expansion and Quintessence phase}\\ [1.3ex]
\toprule[0.7pt]
$k_{6}=\text{odd}$ \; \& \; $k_{4}k_{6}=\text{even}$ \; \& \; $k_{4}+k_{5}B^{k_{6}}<0$ \; \& \; $B<0$ \; \& \; $\vartheta_{2}<B<\vartheta_{1}$ & \tiny{Super-accelerated expansion and Phantom phase}\\ [1.3ex]
\toprule[1.7pt]
\end{tabular}
\end{table*}

On substituting (\ref{ccc8}) into (\ref{v15-1}), the basic reconstruction equation for this case turns out to be
\begin{align}\label{ccc10}
k_{7} \left(\frac{M_{,BB}}{M}-\frac{1}{k_{4}B+k_{5}B^{k_{6}+1}}
\frac{M_{,B}}{M} \right)+\frac{k_{8}}{B}\frac{M_{,B}}{M} -
\frac{k_{9}}{k_{4}B^2+k_{5}B^{k_{6}+2}}=0,
\end{align}
where
\begin{align}\label{ccc11}
k_{7}=& \frac{2(2mg_{0}+1)}{(m+2)\omega_{0}+4(2mg_{0}+1)h_{0}},\nonumber \\
k_{8}=& \left(\frac{2(2m+1)}{(m+2)\omega_{0}+4(2mg_{0}+1)h_{0}}\right)
\nonumber \\ & \times \left(4\frac{f_{0}}{h_{0}}-g_{0}(m+2)\right),\nonumber \\
k_{9}=& \frac{2(2m+1)f_{0}}{(2mg_{0}+1)h^2_{0}}.
\end{align}
Solution to (\ref{ccc10}) is given in terms of the hypergeometric function:
\begin{align}\label{ccc12}
M(B)=\; &k_{16}B^{k_{6}k_{10}} \nonumber \\ & \times \mathbf{hypergeometric}\mathbb{F}\left(k_{10}, k_{11}; k_{12};
 -\frac{k_{5}}{k_{4}}B^{k_{6}}\right) \nonumber \\
+& k_{17}B^{k_{6}k_{13}} \nonumber \\ & \times \mathbf{hypergeometric}\mathbb{F}\left(k_{13}, k_{14}; k_{15};
 -\frac{k_{5}}{k_{4}}B^{k_{6}}\right),
\end{align}
where $k_{16}$ and $k_{17}$ are integration constants, and
\begin{align}\label{ccc13}
k_{10}=&u_{1}+\frac{(k_{7}-k_{8})k_{4}+k_{7}}{2k_{4}k_{6}k_{7}},\nonumber \\
k_{11}=&u_{1}+\frac{(-k_{7}+k_{8})k_{4}+k_{7}}{2k_{4}k_{6}k_{7}},\nonumber \\
k_{12}=&2u_{1}+1,\nonumber \\
k_{13}=&-u_{1}+\frac{(k_{7}-k_{8})k_{4}+k_{7}}{2k_{4}k_{6}k_{7}},\nonumber \\
k_{14}=&-u_{1}+\frac{(-k_{7}+k_{8})k_{4}+k_{7}}{2k_{4}k_{6}k_{7}},\nonumber \\
k_{15}=&-2u_{1}+1,
\end{align}
in which
\begin{align}\label{ccc14}
u_{1}=\frac{\sqrt{(k_{7}-k_{8})^2k^2_{4}+2k_{4}k_{7}(k_{7}-k_{8}+2k_{9})+k^2_{7}   }}{2k_{4}k_{6}k_{7}}.
\end{align}
Clearly, $M(B)$ must be nonzero. In general, it is obtained by putting these conditions: $k_{10}>0$ and $k_{13}>0$ (See `Zeros of hypergeometric function' in ref. \cite{book1}). To find a solution invertible with respect to the scale factor, $B$, one may take one of the integration constants equal to zero (i.e. $k_{16}=0$ or $k_{17}=0$) and then search
for values of the parameters of arguments of hypergeometric function such that it reduces to simple functions or the hypergeometric series terminates. Physics of the problem restricts our freedoms in using the connectional formulas. In several papers, some special forms (without any physical reason) have for simplicity been assumed for the form of the function which connected with hypergeometric functions. But we prefer to set the constants instead of imposing special forms on $M(B)$ for obtaining an invertible $M(B)$ with respect to $B$. In order to this, four options are studied as follows.

\begin{itemize}\label{textcase1}
  \item \noindent\qrmybox{red}{\textbf{Case 1:}}
\end{itemize}
According to the formula
\begin{align}\label{ROF1}
\mathbf{hypergeometric}\mathbb{F}\left(1,1;2;z\right)= -z^{-1} \ln \left( 1-z\right),
\end{align}
which holds for principal branches when $|z|<1$, and by analytic continuation elsewhere \cite{book1,book2}, if we take
      \begin{align}\label{case1-chap-1.1}
        k_{10}=k_{11}=1, \qquad k_{12}=2, \qquad k_{17}=0,
      \end{align}
and set
      \begin{align}\label{case1-chap-1}
        & \left|-\frac{k_{5}}{k_{4}}B^{k_{6}}\right|
        =\left|k_{1}k_{3}B^{k_{6}}\right|<1 \nonumber\\ & \Longrightarrow \quad
        -\left(\frac{1}{k_{1}k_{3}} \right)^{\frac{1}{k_{6}}}<B<\left(\frac{1}{k_{1}k_{3}} \right)^{\frac{1}{k_{6}}},
      \end{align}
      such that in throughout the evolution domain of the scale factor to  be correct, then
      \begin{align}
       M(B) &=\frac{k_{16}}{k_{1}k_{3}} \ln \left[1+k_{1}k_{3}B^{k_{6}} \right], \label{case1-chap-2} \\
       B(\varphi) &=\left(\frac{1}{k_{6}}\left[\frac{\varphi}{C} \right]^{\theta}-\frac{1}{k_{1}k_{3}} \right)^{\frac{1}{k_{6}}},\label{case1-chap-3}
      \end{align}
and
      \begin{align}\label{case1-chap-4}
       \varphi(B)=C\left(\frac{k_{16}}{k_{1}k_{3}} \ln \left[1+k_{1}k_{3}B^{k_{6}} \right
       ]
       \right)^{\frac{1}{\theta}}
      \end{align}
      are obtained. Because the approach and calculations are in the same way like the earlier cases, henceforth we ignore to explain details of calculations.  In view of (\ref{ex}), (\ref{case1-chap-1}), and (\ref{case1-chap-2}) we arrive at the following conditions for the lower bound of the scale factor
      \begin{align}
      & B>\left(-k_{1}k_{3}\right)^{\frac{-k_{2}}{2k_{1}(1+\nu)}},
       \label{case1-chap-5}\\
      & B>-\left(\frac{1}{k_{1}k_{3}} \right)^{\frac{1}{k_{6}}},\label{case1-chap-6}
      \end{align}
and
      \begin{align}\label{case1-chap-7}
      B>\Re \left[\left(\frac{-1}{k_{1}k_{3}} \right)^{\frac{1}{k_{6}}}\right],
      \end{align}
      respectively, which must be held for its amounts in all times of interest. Here, $\Re[...]$ is the real part of $[...]$. Note that we have no worry about the upper and also lower bounds mentioned in (\ref{case1-chap-1}) because the term $(k_{2}k_{3})^{-1/k_{6}}$ can be taken so large in comparison with the scale factor by setting the constant parameters. The present absolute amount of the average scale factor is one, so, the condition $(k_{1}k_{3})^{-1/k_{6}}\gg 1$ may be established. The importance of the three bounds in (\ref{case1-chap-5}), (\ref{case1-chap-6}), and (\ref{case1-chap-7}) is that when the positive-valued scale factor is considered, then
\begin{align}\label{supremum}
B> \textbf{Supremum}\left\{\left(-k_{1}k_{3}\right)^{\frac{-k_{2}}{2k_{1}(1+\nu)}},
\; -\left(\frac{1}{k_{1}k_{3}} \right)^{\frac{1}{k_{6}}}, \Re \left[\left(\frac{-1}{k_{1}k_{3}} \right)^{\frac{1}{k_{6}}}\right] \right\},
\end{align}
must be adopted, and for the negative-valued scale factor, the condition
\begin{align}\label{infimum}
B< \textbf{Infimum} \left\{\left(-k_{1}k_{3}\right)^{\frac{-k_{2}}{2k_{1}(1+\nu)}},
\; -\left(\frac{1}{k_{1}k_{3}} \right)^{\frac{1}{k_{6}}}, \Re \left[\left(\frac{-1}{k_{1}k_{3}} \right)^{\frac{1}{k_{6}}}\right] \right\},
\end{align}
would be the case. Limpidly, both positive and negative sings are admissible for the scale factor, but, for simplicity, let us consider the positive one. Therefore, (\ref{case1-chap-6}) is removed automatically and (\ref{case1-chap-1}) reduces to $0<B<(k_{1}k_{3})^{-1/k_{6}}$ or equivalently $0<k_{1}k_{3}B^{k_{6}}<1$ and consequently, $0<\ln \left[1+k_{1}k_{3}B^{k_{6}} \right]<0.69$. According to $\mathfrak{B}[B,\alpha_{0}=0; \varphi(B)]<0$ and $\mathfrak{B}[\varphi, \beta_{0}=0; B(\varphi)]<0$, one has
      \begin{align}\label{case1-chap-10}
      \theta <0
      \end{align}
      and
      \begin{align}\label{case1-chap-11}
      \frac{k_{1}k_{3}}{k_{6}}\left(\frac{\varphi}{C} \right)^{\theta}>1,
      \end{align}
      respectively. The conditions (\ref{chaharshart1})-(\ref{chaharshart2}) also reduce to (\ref{case1-chap-10}). From (\ref{case1-chap-3}), (\ref{case1-chap-11}), and Table (\ref{table1-dec1}) it is argued that this case of the solution is maintained only for decelerated or accelerated era, depending upon the amounts of $k_{1}$ and $k_{2}$.\\
      Here, the form of the potential in terms of the scale factor and also scalar field are a little inconvenient and complicated such as \(F(R)\)-gravity:
\begin{align}
       V(B)=&C^2\left(c_{14}B^{\frac{-2k_{1}(1+\nu)}{k_{2}}}+\frac{k_{3}}{k_{1}} \right)^{\frac{1}{(1+\nu)}} \nonumber \\ & \times \left(\frac{k_{16}}{k_{1}k_{3}}\ln\left[1+k_{1}k_{3}B^{k_{6}}\right]\right)
^{\frac{2}{\theta}} \nonumber \\
&\times \left\{k_{18}+\frac{B^{k_{6}}}{\left(1+k_{1}k_{3}B^{k_{6}}\right)\ln \left[1+k_{1}k_{3}B^{k_{6}} \right]}\right. \nonumber \\ & \left.\left(k_{19}+k_{20}\frac{B^{k_{6}}}{\left(1+k_{1}k_{3}B^{k_{6}}\right)
\ln \left[1+k_{1}k_{3}B^{k_{6}} \right]} \right) \right\},\label{case1-chap-12}\\
V(\varphi)=&C^2 \left(c_{14}\left(\frac{1}{k_{6}}\left[\frac{\varphi}{C} \right]^{\theta}-\frac{1}{k_{1}k_{3}} \right)^{\frac{-2k_{1}(1+\nu)}{k_{2}k_{6}}}+\frac{k_{3}}{k_{1}} \right)^{\frac{1}{(1+\nu)}} \nonumber \\ & \times \left(\frac{k_{16}}{k_{1}k_{3}}\ln \left[ \frac{k_{1}k_{3}}{k_{6}}\left(\frac{\varphi}{C} \right)^{\theta}\right] \right)^{\frac{2}{\theta}} \nonumber \\
&\times \left \{k_{18}+\frac{\left(\frac{1}{k_{6}}\left(\frac{\varphi}{C} \right)^{\theta}-\frac{1}{k_{1}k_{3}}\right)}
{\left(\frac{k_{1}k_{3}}{k_{6}}\left(\frac{\varphi}{C} \right)^{\theta}  \right)\ln \left[\frac{k_{1}k_{3}}{k_{6}}\left(\frac{\varphi}{C} \right)^{\theta} \right]} \right. \nonumber \\ & \left. \left(k_{19}+k_{20}\frac{\left(\frac{1}{k_{6}}\left(\frac{\varphi}{C} \right)^{\theta}-\frac{1}{k_{1}k_{3}}\right)}
{\left(\frac{k_{1}k_{3}}{k_{6}}\left(\frac{\varphi}{C} \right)^{\theta}  \right)\ln \left[\frac{k_{1}k_{3}}{k_{6}}\left(\frac{\varphi}{C} \right)^{\theta} \right]} \right) \right\},\label{case1-chap-13}
      \end{align}
where
\begin{align}\label{case1-chap-14}
&k_{18}=2(2m+1)f^2_{0}, \quad k_{19}=\frac{2}{\theta}h_{0}k_{1}k_{3}k_{6}(2mg_{0}+1),\nonumber \\& k_{20}=\frac{-\omega_{0}}{2\theta^2}k^2_{1}k^2_{3}.
\end{align}
However, the explicit expressions for the potential in both versions are cumbersome and long, but it is worth to note that because $0<k_{1}k_{3}B^{k_{6}}<1$, therefore $\ln \left[ 1+k_{1}k_{3}B^{k_{6}}\right]$ and consequently $\ln \left[\frac{k_{1}k_{3}}{k_{6}}\left(\frac{\varphi}{C} \right)^{\theta} \right]$ --- because of $1+k_{1}k_{3}B^{k_{6}}=\frac{k_{1}k_{3}}{k_{6}}\left(\frac{\varphi}{C} \right)^{\theta}$ --- can be expressed via Taylor series:
\begin{align}\label{case1-chap-15}
\ln\left[1+x \right]=\sum_{n=1}^{\infty}\frac{(-1)^{n+1}}{n}x^n \; ; \quad -1<x\leq 1.
\end{align}
It is also needless to run this series from one to infinity, because as mentioned earlier, the scalar field goes down, so we can use a reasonable approximation (for example $\mathcal{O}(\varphi^{2\theta})$). Furthermore, we kept all constants without any selections for their values, otherwise, it may be written in the more convenient form:
\begin{align}\label{case1-chap-16}
V(\varphi)\simeq & V_{0}\varphi^2 \left(\varphi^{\frac{3}{2}\left|\theta \right|}+\varphi^{\frac{5}{2}\left|\theta \right|} \right)
\nonumber \\ &=V_{0} \varphi^{\frac{7}{2}\left|\theta \right|} \left(1+\varphi^{\left|\theta \right|} \right),
\end{align}
in which we have taken $\nu=1$ and $c_{14}\gg k_{1}/k_{3}$, and $V_{0}$ is a constant. For different amounts of $\left|\theta \right|$, different orders of the scalar potential are available.\\
The potential conditions do not present straightforward constraints for the parameters in (\ref{case1-chap-12}) and (\ref{case1-chap-13}), because of the form of potential, only in data analysis it would be beneficial.\\
For the potential (\ref{case1-chap-16}), the function $\Gamma$ would be complicated when there is no selection for the amount of $\theta$. But it is not hard to specify the type of the potential for a given $\theta$, for example $\theta=2$ acquires `Thawing' type.

\begin{itemize}\label{textcase2}
  \item \noindent\qrmybox{red}{\textbf{Case 2:}}
\end{itemize}
In this case, we want to utilize the formula
\begin{align}\label{ROF2}
\mathbf{hypergeometric}\mathbb{F}\left(\frac{1}{2},1;\frac{3}{2};-z^2\right)= z^{-1} \arctan(z),
\end{align}
which holds for principal branches when $|z|<1$, and by analytic continuation elsewhere \cite{book1,book2}. Hence by demanding
\begin{align}\label{case1-chap-17}
k_{10}=\frac{1}{2}, \quad k_{11}=1, \quad k_{12}=\frac{3}{2}, \quad k_{17}=0,
\end{align}
and
\begin{align}\label{case1-chap-18}
&\left| \sqrt{k_{1}k_{3}B^{k_{6}}}\right|<1 \quad
\Longrightarrow \quad 0<B<\left(k_{1}k_{3}\right)^{\frac{-1}{k_{6}}},
\end{align}
we arrive at
\begin{align}
M(B)&=\frac{k_{16}}{\sqrt{k_{1}k_{3}}}\arctan\left[\sqrt{k_{1}k_{3}
B^{k_{6}}}\right],\label{case1-chap-19} \\
\varphi(B)&=C\left(\frac{k_{16}}{\sqrt{k_{1}k_{3}}}\arctan\left[\sqrt{k_{1}k_{3}
B^{k_{6}}}\right] \right)^{\frac{1}{\theta}},\label{case1-chap-20}
\end{align}
and
\begin{align}\label{case1-chap-21}
B(\varphi)=\left(\frac{1}{k_{1}k_{3}}\tan^2 \left[\frac{\sqrt{k_{1}k_{3}}}{k_{16}}\left(\frac{\varphi}{C} \right)^{\theta} \right] \right)^{\frac{1}{k_{6}}}.
\end{align}
In view of (\ref{case1-chap-18}), there is an upper bound for the scale factor which is not so satisfying. Hence, without loss of generality, let us take $k_{3}$ to be too small, roughly near zero. Moreover, this ans\"{a}tz would be so helpful for extracting a suitable form for the scalar potential. According to (\ref{case1-chap-21}) and the domain of the tangent function, by putting the condition
\begin{align}\label{case1-chap-22}
\left|\frac{\sqrt{k_{1}k_{3}}}{k_{16}}\left(\frac{\varphi}{C} \right)^{\theta} \right|< \frac{\pi}{2},
\end{align}
we do not have the singular points as the scale factor evolves. The conditions (\ref{chaharshart1})-(\ref{chaharshart2}), $\mathfrak{B}[\varphi, \beta_{0}=0; B(\varphi)]<0$, and $\mathfrak{B}[B, \alpha_{0}=0; \varphi(B)]<0$ lead to the following conditions:
\begin{align}\label{case1-chap-23}
\theta <0, \qquad \frac{\varphi}{\left|\varphi \right|}\frac{C}{\left|C \right|}>0,
\end{align}
\begin{equation}\label{case1-chap-24}\begin{split}
\left\{
  \begin{array}{ll}
    \text{1: } \; k_{16}>0  \; \Longleftrightarrow \;
0<\frac{\sqrt{k_{1}k_{3}}}{k_{16}}\left(\frac{\varphi}{C} \right)^{\theta} < \frac{\pi}{2}, & \\ \\
    \text{2: } \; k_{16}<0 \; \Longleftrightarrow \;
\frac{-\pi}{2}<\frac{\sqrt{k_{1}k_{3}}}{k_{16}}\left(\frac{\varphi}{C} \right)^{\theta} <0 , &
  \end{array}
\right.
\end{split}\end{equation}
Therefore, (\ref{case1-chap-22}) reduces to (\ref{case1-chap-24}). Using (\ref{case1-chap-21}) and (\ref{case1-chap-24}) and comparing with Table (\ref{table1-dec1}), it is realized that this case of the solution is in accelerated or decelerated era, depending upon the values of $k_{1}$ and $k_{2}$.\\
The form of the scalar potential in terms of the scale factor and scalar field become complicated:
\begin{align}\label{case1-chap-25}
V(B)=&\left(c_{14}B^{\frac{-2k_{1}(1+\nu)}{k_{2}}}+\frac{k_{3}}{k_{1}} \right)^{\frac{1}{(1+\nu)}} \nonumber \\ & \times \left(\frac{k_{16}}{\sqrt{k_{1}k_{3}}}\arctan \left[ \sqrt{k_{1}k_{3}B^{k_{6}}}\right] \right)^{\frac{2}{\theta}} \nonumber \\
&\times \biggl\{2(2m+1)f_{0}C^2 \nonumber \\ &+\frac{C^2 k_{6}h_{0}(2mg_{0}+1)\sqrt{k_{1}k_{3}B^{k_{6}}}}{\theta \left(1+k_{1}k_{3}B^{k_{6}} \right)\arctan \left[\sqrt{k_{1}k_{3}B^{k_{6}}} \right]} \nonumber \\ & -\frac{C^2\omega_{0}k_{1}k_{3}k^2_{6}B^{k_{6}}}{8\theta^2 \left(1+k_{1}k_{3}B^{k_{6}} \right)^2 \arctan^{2} \left[\sqrt{k_{1}k_{3}B^{k_{6}}} \right]}  \biggl\},
\end{align}
and
\begin{align}\label{case1-chap-26}
V(\varphi)&=\left(\frac{c_{14}k_{1}k_{3}}
{\tan^2\left[\frac{\sqrt{k_{1}k_{3}}}{k_{16}}
\left(\frac{\varphi}{C}\right)^{\theta} \right]}
+\frac{k_{3}}{k_{1}} \right)^{\frac{1}{(1+\nu)}} \left(\frac{\varphi}{C} \right)^{2} \nonumber \\
&\times \biggl\{2(2m+1)f_{0}C^2 + \frac{C^2k_{6}k_{16}h_{0}(2mg_{0}+1)}{\theta \sqrt{k_{1}k_{3}}}
\nonumber \\ & \times \frac{\tan\left[\frac{\sqrt{k_{1}k_{3}}}{k_{16}}
\left(\frac{\varphi}{C}\right)^{\theta} \right]}{1+\tan^2\left[\frac{\sqrt{k_{1}k_{3}}}{k_{16}}
\left(\frac{\varphi}{C}\right)^{\theta} \right]}\left(\frac{\varphi}{C} \right)^{-\theta}
\nonumber \\&- \frac{C^2 \omega_{0}k^2_{6}k^2_{16}}{8k_{1}k_{3}\theta^2}
\frac{\tan^2\left[\frac{\sqrt{k_{1}k_{3}}}{k_{16}}
\left(\frac{\varphi}{C}\right)^{\theta} \right]}{\left(1+\tan^2\left[\frac{\sqrt{k_{1}k_{3}}}{k_{16}}
\left(\frac{\varphi}{C}\right)^{\theta} \right] \right)^2}
\left( \frac{\varphi}{C}\right)^{-2\theta} \biggl\},
\end{align}
or put differently
\begin{align}\label{case1-chap-27}
V(\varphi)&=\left(\frac{c_{14}k_{1}k_{3}}
{\tan^2\left[\frac{\sqrt{k_{1}k_{3}}}{k_{16}}
\left(\frac{\varphi}{C}\right)^{\theta} \right]}
+\frac{k_{3}}{k_{1}} \right)^{\frac{1}{(1+\nu)}} \left(\frac{\varphi}{C} \right)^{2} \nonumber \\
&\times \biggl\{2(2m+1)f_{0}C^2  \nonumber \\ &+ \frac{C^2k_{6}k_{16}h_{0}(2mg_{0}+1)}{2\theta \sqrt{k_{1}k_{3}}}
\sin\left[2\frac{\sqrt{k_{1}k_{3}}}{k_{16}}
\left(\frac{\varphi}{C}\right)^{\theta} \right]\left(\frac{\varphi}{C} \right)^{-\theta}
\nonumber \\&- \frac{C^2 \omega_{0}k^2_{6}k^2_{16}}{32k_{1}k_{3}\theta^2}
\sin^2\left[2\frac{\sqrt{k_{1}k_{3}}}{k_{16}}
\left(\frac{\varphi}{C}\right)^{\theta} \right]
\left( \frac{\varphi}{C}\right)^{-2\theta} \biggl\}.
\end{align}
Propitiously, under the assumption we are using (i.e. $k_{3}\lll$), this form of the potential is now simplified enormously as:
\begin{align}\label{case1-chap-28}
V(\varphi)\simeq V_{0ch_{2}}\left(\frac{\varphi}{C}\right)^{2}
\left(\beta_{c1}\left(\frac{\varphi}{C} \right)^{\frac{-\theta}{1+\nu}} +\beta_{c2}\left(\frac{\varphi}{C} \right)^{\left(\frac{\nu}{1+\nu} \right)\theta} \right),
\end{align}
where
\begin{align}\label{case1-chap-29}
V_{0ch2}&=2(2m+1)f_{0}C^2 + \frac{C^2k_{6}h_{0}(2mg_{0}+1)}{\theta}- \frac{C^2 \omega_{0}k^2_{6}}{8\theta^2}, \nonumber \\
\beta_{c1}&=c_{14}k_{14}, \nonumber \\
\beta_{c2}&=\frac{k_{3}c^{\frac{-\nu}{1+\nu}}_{14}
k^{\frac{1}{1+\nu}}_{14}}{k_{1}k_{16}(1+\nu)}.
\end{align}
Note that we have not neglected the terms containing $k_{3}$, rather we have used it to approximate the `Sinus' and `Tangent' functions up to their first terms. Thanks to this reasonable assumption, in practice, the well-known potentials which have the form as $V=V_{0}\varphi^{2n}$, where $n$ is a constant, may be extracted:
\begin{align}\label{case1-chap-30}
V(\varphi)\simeq \tilde{V}_{och2}\varphi^{2n}; \quad \text{where}
\quad \tilde{V}_{och2}=\frac{V_{och2}\beta_{c1}}{C^{2n}},
\end{align}
provided that we choose
\begin{align}\label{case1-chap-31}
\theta=-2^n, \qquad \nu=\frac{2}{2^n -2}.
\end{align}
Limpidly, here we cannot accept the range $-0.2< \nu < 1.8$, rather it must reduce to $0< \nu < 1.8$ and hence it consequently leads to $n>1.637$. Consequently, we obtain $\Gamma<1$ (`Thawing').

  \begin{itemize}\label{textcase3}
  \item \noindent\qrmybox{red}{\textbf{Case 3:}}
  \end{itemize}
In this case, we proceed with the formula
\begin{align}\label{ROF3}
\mathbf{hypergeometric}\mathbb{F}
\left(\frac{1}{2},\frac{1}{2};\frac{3}{2};-z^2\right)= z^{-1} \ln \left( z+\sqrt{1+z^2}\right),
\end{align}
which holds for principal branches when $|z|<1$, and by analytic continuation elsewhere \cite{book1,book2}. In order to follow it, we set
\begin{align}\label{case1-chap-32}
k_{10}=k_{11}=\frac{1}{2}, \quad k_{12}=\frac{3}{2}, \quad k_{17}=0,
\end{align}
and
\begin{align}\label{case1-chap-32.1}
\left| \sqrt{k_{1}k_{3}B^{k_{6}}}\right|<1 \quad \Longrightarrow \quad 0<B<\left(k_{1}k_{3}\right)^{\frac{-1}{k_{6}}}.
\end{align}
Therefore, the forms of $M(B)$, $\varphi(B)$, and $B(\varphi)$ turn out to be
\begin{align}
M(B)=&\frac{k_{16}}{\sqrt{k_{1}k_{3}}} \ln \left[\sqrt{k_{1}k_{3}B^{k_{6}}}+\sqrt{1+k_{1}k_{3}B^{k_{6}}} \right] \nonumber \\
=&\frac{k_{16}}{\sqrt{k_{1}k_{3}}} \text{arcsinh}\left[\sqrt{k_{1}k_{3}B^{k_{6}}} \right],\label{case1-chap-33} \\
\varphi(B)=&C \left(\frac{k_{16}}{\sqrt{k_{1}k_{3}}} \ln \left[\sqrt{k_{1}k_{3}B^{k_{6}}}+\sqrt{1+k_{1}k_{3}B^{k_{6}}} \right] \right)^{\frac{1}{\theta}} \nonumber \\
=&\left(\frac{k_{16}}{\sqrt{k_{1}k_{3}}}
\text{arcsinh}\left[\sqrt{k_{1}k_{3}B^{k_{6}}} \right]\right)^{\frac{1}{\theta}},\label{case1-chap-34}
\end{align}
and
\begin{align}\label{case1-chap-35}
B(\varphi)=\left(\frac{1}{k_{1}k_{3}}
\sinh^2\left[\frac{\sqrt{k_{1}k_{3}}}{k_{16}}\left(\frac{\varphi}{C} \right)^{\theta}\right] \right)^{\frac{1}{k_{6}}},
\end{align}
respectively. Clearly, such as `Case 2', we have no worry about the upper bound of the scale factor, for it may be removed by setting the constants, for instance assuming $k_{3}\lll$. The conditions (\ref{chaharshart1})-(\ref{chaharshart2}) and $\mathfrak{B}[B,\alpha_{0}=0; \varphi(B)]<0$ give
\begin{align}\label{case1-chap-36}
\theta<0,
\end{align}
and the conditions
\begin{align}\label{case1-chap-37}
\left\{
  \begin{array}{ll}
    1:\; k_{16}>0 \Longleftrightarrow \frac{\sqrt{k_{1}k_{3}}}{k_{16}}\left( \frac{\varphi}{C}\right)^{\theta}>0, & \\ \\
    2:\; k_{16}<0 \Longleftrightarrow \frac{\sqrt{k_{1}k_{3}}}{k_{16}}\left( \frac{\varphi}{C}\right)^{\theta}<0, &  \\ \\
    3:\; \frac{\varphi}{\left|\varphi\right|}\frac{C}{\left|C\right|}>0, &
  \end{array}
\right.
\end{align}
are emerged by $\mathfrak{B}[\varphi, \beta_{0}=0; B(\varphi)]<0$. Now, we return to the form of the scale factor. As is observed, the form of the scale factor is interesting. For comparison this form of the scale factor with the scale factor of the standard cosmological model, let us review the standard model of the expanding universe, filled with nonrelativistic matter and using the $\Lambda\text{-term}$, dark energy, as the source of acceleration of cosmological expansion. The metric of the standard model is that of the
space-flat FRW universe as $\mathrm{d}s^2=\mathrm{d}t^2-a^2(t)[\mathrm{d}x^2+\mathrm{d}y^2+\mathrm{d}z^2]$. The scale factor $a(t)$ is determined from the Friedmann equation, which may be written in the form:
\begin{align}\label{case1-chap-38.11}
\left(\frac{\dot{a}(t)}{a(t)} \right)^2=H^2_{0} \left[\Omega_{m}\left(\frac{a(t_{0})}{a(t)} \right)^3+\Omega_{\Lambda} \right],
\end{align}
where $H_{0}$ and $a(t_{0})$ are the current amounts of the Hubble parameter and scale factor, and $\Omega_{m}$ and $\Omega_{\Lambda}$ are the current values of the matter and cosmological constant densities, respectively. It has been demonstrated in ref. \cite{akhar1231} that a solution to (\ref{case1-chap-38.11}) is
\begin{align}\label{case1-chap-38}
a(t)=\left(\frac{\Omega_{m}}{\Omega_{\Lambda}} \right)^{\frac{1}{3}}
\left( \sinh\left[\frac{3}{2}\sqrt{\Omega_{\Lambda}}H_{0}t \right]\right)^{\frac{2}{3}}.
\end{align}
It is an established common knowledge in gravity that this form of the scale factor is successful in explaining the elaborations of important events such as phase crossing from quintessence to phantom, and shifting from decelerated to accelerated expansion. Furthermore, the two limiting cases of this scale factor namely $H_{0}t\ll 1$ and $H_{0}t\gg 1$ which lead to $a(t)=(9\Omega_{m}/4)^{1/3}(H_{0}t)^{2/3}$ and $a(t)=(\Omega_{m}/4\Omega_{\Lambda})^{1/3}
\exp\left[\sqrt{\Omega_{\Lambda}}H_{0}t \right]$, respectively, are also notable as the former one corresponds to the scale factor in the stage of matter domination and the later one corresponds to the scale factor of a universe with a de Sitter expansion law. The interesting point here is that the scale factor (\ref{case1-chap-35}) to the power of $(m+2)/3$, (Note that $a_{ave.}=B^{(m+2)/3}$), recovers (\ref{case1-chap-38}) by taking
\begin{align}\label{case1-chap-39}
&\theta=-1,\quad \frac{m+2}{k_{6}}=1, \quad \varphi(t)=\frac{\varphi_{0}}{t}, \nonumber \\ &k_{1}k_{3}=\left( \frac{\Omega_{m}}{\Omega_{\Lambda}}\right)^{\frac{1}{3}}, \quad
\frac{C}{\varphi_{0}k_{16}}=\frac{3}{2}H_{0} \Omega^{\frac{1}{6}}_{m} \Omega^{\frac{1}{3}}_{\Lambda},
\end{align}
where $\varphi_{0}$ is a positive constant. It must be noted that these assumptions are compatible with the conditions obtained by the conditions $\mathfrak{B}[B,\alpha_{0}=0; \varphi(B)]<0$ and $\mathfrak{B}[\varphi, \beta_{0}=0; B(\varphi)]<0$.\\
This case such as the previous cases produces uncomfortable forms for the potential in terms of the scale factor and also the scalar field:
\begin{align}
V(B)=&C^2\left(c_{14}B^{\frac{-2k_{1}(1+\nu)}{k_{2}}}+\frac{k_{3}}{k_{1}} \right)^{\frac{1}{(1+\nu)}} \nonumber \\& \times \left(\frac{k_{16}}{\sqrt{k_{1}k_{3}}} \text{arcsinh}\left[\sqrt{k_{1}k_{3}B^{k_{6}}} \right]\right)^{\frac{2}{\theta}} \nonumber \\
\times& \biggl\{2(2m+1)f_{0} \nonumber \\ &+\frac{h_{0}k_{6}(2mg_{0}+1)
\sqrt{k_{1}k_{3}B^{k_{6}}}}{\theta \sqrt{1+k_{1}k_{3}B^{k_{6}}}
\text{arcsinh}\left[\sqrt{k_{1}k_{3}B^{k_{6}}} \right]} \nonumber \\ & -\frac{\omega_{0}k^2_{6}k_{1}k_{3}B^{k_{6}}}{8\theta^2 \left( 1+k_{1}k_{3}B^{k_{6}}\right)\text{arcsinh}^2\left[\sqrt{k_{1}k_{3}B^{k_{6}}} \right]}\biggl\}\label{case1-chap-40}
\end{align}
\begin{align}
V(\varphi)=&\varphi^{2} \left(c_{14}k_{1}k_{3}\sinh^{-2}\left[\frac{\sqrt{k_{1}k_{3}}}{k_{16}}\left( \frac{\varphi}{C}\right)^{\theta} \right]+\frac{k_{3}}{k_{1}} \right)^{\frac{1}{(1+\nu)}} \nonumber \\
&\times\biggl\{2(2m+1)f_{0}+\frac{h_{0}k_{6}k_{16}(2mg_{0}+1)}{\theta \sqrt{k_{1}k_{3}}}\left(\frac{\varphi}{C} \right)^{-\theta} \nonumber \\ & \times \tanh\left[\frac{\sqrt{k_{1}k_{3}}}{k_{16}}\left( \frac{\varphi}{C}\right)^{\theta} \right]
\nonumber \\&-\frac{\omega_{0}k^2_{6}k^2_{16}}{8\theta^2 k_{1}k_{3}}\left( \frac{\varphi}{C}\right)^{-2\theta}
\tanh^2\left[\frac{\sqrt{k_{1}k_{3}}}{k_{16}}\left( \frac{\varphi}{C}\right)^{\theta} \right] \biggl\}.\label{case1-chap-41}
\end{align}
Taking $c_{14}=1/k^2_{1}$, (\ref{case1-chap-41}) yields
\begin{align}\label{case1-chap-42}
V(\varphi)=&\varphi^{2} \left(\frac{k_{3}}{k_{1}}\right)^{\frac{1}{(1+\nu)}}
\coth^{\frac{2}{(1+\nu)}}\left[\frac{\sqrt{k_{1}k_{3}}}{k_{16}}\left( \frac{\varphi}{C}\right)^{\theta} \right] \nonumber \\
&\times\biggl\{2(2m+1)f_{0}+\frac{h_{0}k_{6}k_{16}(2mg_{0}+1)}{\theta \sqrt{k_{1}k_{3}}}\left(\frac{\varphi}{C} \right)^{-\theta} \nonumber \\ & \times \tanh\left[\frac{\sqrt{k_{1}k_{3}}}{k_{16}}\left( \frac{\varphi}{C}\right)^{\theta} \right]
\nonumber \\&-\frac{\omega_{0}k^2_{6}k^2_{16}}{8\theta^2 k_{1}k_{3}}\left( \frac{\varphi}{C}\right)^{-2\theta}
\tanh^2\left[\frac{\sqrt{k_{1}k_{3}}}{k_{16}}\left( \frac{\varphi}{C}\right)^{\theta} \right] \biggl\}.
\end{align}
Such as `case 2', since the physics of the problem implies that $k_{3}$ to be very small --- so that we have $(\sqrt{k_{1}k_{3}}/k_{16})(\varphi/C)^{\theta}\lll $ ---, the form of the potential would be the case of interest as:
\begin{align}\label{case1-chap-43}
V(\varphi)=V_{0ch3}\varphi^{2+\frac{2\left|\theta\right|}{(1+\nu)}},
\end{align}
where
\begin{align}\label{case1-chap-44}
V_{0ch3}=\left(\frac{k_{16}}{k_{1}C^{\left|\theta\right|}}\right)
^{\frac{2}{(1+\nu)}} \left(2f_{0}(2m+1)
+\frac{h_{0}k_{6}(2mg_{0}+1)}{\theta}-\frac{\omega_{0}k^2_{6}}{8\theta^2} \right).
\end{align}
Another limiting case namely $(\sqrt{k_{1}k_{3}}/k_{16})(\varphi/C)^{\theta}\ggg$, where we have assumed that $k_{1}\sim 1/k_{3}$ (To cancel very small value of $k_{3}$) and $(C^{\left|\theta\right|}/k_{16})\ggg $, leads to
\begin{align}\label{case1-chap-45}
V(\varphi)=\varphi^2 \left(\alpha_{c31}+\alpha_{c32}\varphi^{\left|\theta\right|}
+\alpha_{c33}\varphi^{2\left|\theta\right|} \right),
\end{align}
where
\begin{align}
\alpha_{c31}&=2f_{0}(2m+1)\left(\frac{k_{3}}{k_{1}} \right)^{\frac{1}{(1+\nu)}},\label{case1-chap-46}\\
\alpha_{c32}&=\frac{h_{0}k_{6}k_{16}k^{\frac{(1-\nu)}{2(1+\nu)}}_{3}
(2mg_{0}+1)}{\theta C^{\left|\theta\right|}k^{\frac{(3+\nu)}{2(\nu+1)}}_{1}},\label{case1-chap-47}\\
\alpha_{c33}&=\frac{-\omega_{0}k^2_{6}k^2_{16}}{8\theta^2 C^{2\left|\theta\right|}k^{\frac{(2+\nu)}{(1+\nu)}}_{1}
k^{\frac{\nu}{(1+\nu)}}_{3}}.\label{case1-chap-48}
\end{align}
This form of the potential may be written in a good closed form as
\begin{align}\label{case1-chap-49}
V(\varphi)=\alpha_{c32}\varphi^2 \left(\sqrt{\frac{\alpha_{c31}}{\alpha_{c32}}}+\varphi^{\left|\theta\right|} \right)^2,
\end{align}
where we have set $\alpha_{c32}=2\sqrt{\alpha_{c31}\alpha_{c32}}$. It is worth noting that unlike the first limiting case, here the parameter $\nu$ does not have an effect on the power of the scalar field. This means that in the de Sitter universe, both generalized and standard Chaplygin gases lead to the same potential up to a constant factor.\\
Let us carry another challenging discussion out: Suppose that in (\ref{case1-chap-49}), the constant $\sqrt{\alpha_{c31}/\alpha_{c32}}$ is negligible in comparison with $\varphi^{\left|\theta\right|}$ in the throughout of the evolution range of the scalar field. Now, compare it with the first limiting case:
\begin{equation}\label{case1-chap-50}\begin{split}
\left\{
\begin{array}{ll}
 V(\varphi)_{1}\thicksim \varphi^{2+\frac{2\left|\theta\right|}{(1+\nu)}} \; \quad &{} \; \text{The first limiting case}\\&{} \text{(which corresponds to matter} \\&{} \text{dominated era);}\\ \\
V(\varphi)_{2}\thicksim \varphi^{2+2\left|\theta\right|} \; &{} \text{The second limiting case} \\&{} \text{(which corresponds to de Sitter)} \\ &{} \text{with a further assumption at} \\ &{} \text{all times of interest:} \\&{} \sqrt{\alpha_{c31}/\alpha_{c32}}\lll \varphi^{2\left|\theta\right|}
\end{array}
\right.
\end{split}\end{equation}
From the unification vantage point, if the last two stages of the universe are described by a unique form of the potential, then, `perhaps', we may allow a time dependence for the parameter $\nu$ (which we hypothesized it is a constant). Therefore, if it is an admissible postulate, then we can conclude that $\nu(t)$ drops with time (from an initial constant value which is in $-0.2< \nu < 1.8$ or maybe in $|\nu|\lesssim 0.05$, to zero). As mentioned earlier, $\nu$ is very close to zero when describing the cosmological dark sector is demanded via the generalized Chaplygin gas \cite{darksector}, hence, our finding coincides with this. Therefore, the generalized Chaplygin gas model (which was $P=-\sigma_{2} / \rho^{\nu}(t)$) may be developed as $P=-\sigma_{2} / \rho^{\nu(t)}(t)$ (i.e. $\nu=const. \longrightarrow \nu(t)$).\\
For both limiting forms of the potential, we have $\Gamma<1$ (i.e. `Thawing' type).

\begin{itemize}\label{textcase4}
  \item \noindent\qrmybox{red}{\textbf{Case 4:}}
\end{itemize}
This case is the simplest one among the cases of interest. To utilize the relation
\begin{align}\label{ROF4}
\mathbf{hypergeometric}\mathbb{F}
\left(a,b;b;z\right)= (1-z)^{-a},
\end{align}
which holds for principal branches when $|z|<1$, and by analytic continuation elsewhere \cite{book1,book2}, we make use of the following assumptions
\begin{align}\label{case1-chap-51}
k_{11}=k_{12}, \quad k_{10}>0, \quad k_{17}=0,
\end{align}
and
\begin{align}\label{case1-chap-52}
& \left|-\frac{k_{5}}{k_{4}}B^{k_{6}}\right|
        =\left|k_{2}k_{3}B^{k_{6}}\right|<1 \nonumber \\& \Longrightarrow
        -\left(\frac{1}{k_{2}k_{3}} \right)^{\frac{1}{k_{6}}}<B<\left(\frac{1}{k_{2}k_{3}} \right)^{\frac{1}{k_{6}}}.
\end{align}
Note that the second condition $k_{10}>0$ was set earlier (See the first line after eq. (\ref{ccc14})). For simplicity, let us restrict ourselves to the positive values of the scale factor. Here, like three cases studied, $k_{3}$ must be very small. Carrying the simple calculations out, one has
\begin{align}
M(B) &=k_{16}\left(1+k_{1}k_{3}B^{k_{6}} \right)^{-k_{10}}, \label{case1-chap-53} \\
\varphi(B) &=C \left[k_{16}\left(1+k_{1}k_{3}B^{k_{6}} \right)^{-k_{10}} \right]^{\frac{1}{\theta}},\label{case1-chap-54}
\end{align}
and
\begin{align}\label{case1-chap-55}
B(\varphi)=\left(\frac{1}{k_{1}k_{3}}\left[\left( \frac{1}{k_{16}}\left[ \frac{\varphi}{C}\right]^{\theta}\right)^{\frac{-1}{k_{10}}}-1 \right] \right)^{\frac{1}{k_{6}}}.
\end{align}
Unlike the other cases, here, the condition $\mathfrak{B}[\varphi,\beta_{0}=0; B(\varphi)]<0$ implies that the parameter $\theta$ must be positive:
\begin{align}\label{case1-chap-56}
\frac{-k_{1}k_{3}k_{6}k_{10}B^{k_{6}}}{\theta \left(1+k_{1}k_{3}B^{k_{6}} \right)}<0 \quad \Longrightarrow \quad \theta >0,
\end{align}
and the conditions (\ref{chaharshart1})-(\ref{chaharshart2}) and $\mathfrak{B}[B,\alpha_{0}=0; \varphi(B)]<0$ yield
\begin{align}\label{11case1-chap-57}
\frac{\varphi}{\left|\varphi \right|}\frac{C}{\left|C\right|}>0,
\end{align}
and
\begin{align}\label{case1-chap-57}
\frac{1}{k_{16}}\left(\frac{\varphi}{C} \right)^{\theta}<1.
\end{align}
The condition (\ref{case1-chap-57}) is not an intense upper bound because we have three degrees of freedom that can help us stretch the wide of bound for the scalar field.\\
The forms of the potential in terms of the scale factor and scalar field become:
\begin{align}
V(B)=&\left(c_{14}B^{\frac{-2k_{1}(1+\nu)}{k_{2}}}+\frac{k_{3}}{k_{1}} \right)^{\frac{1}{(1+\nu)}} \nonumber \\ & \times \left[k_{16}\left(1+k_{1}k_{3}B^{k_{6}} \right)^{-k_{10}} \right]^{\frac{2}{\theta}} \biggl\{2(2m+1)f_{0}C^2 \nonumber \\ &-\frac{2(2mg_{0}+1)h_{0}C^2 k_{1}k_{3}k_{6}k_{10}B^{k_{6}}}{\theta \left(1+k_{1}k_{3}B^{k_{6}}\right)} \nonumber \\ &- \frac{\omega_{0}}{2}\left(\frac{Ck_{1}k_{3}k_{6}k_{10}B^{k_{6}}}{\theta \left( 1+k_{1}k_{3}B^{k_{6}}\right)}\right)^2  \biggl\},\label{case1-chap-58}\\
V(\varphi)&=\varphi^2 \nonumber \\ & \times \left[c_{14}k_{1}k_{3}\left(\left[ \frac{1}{k_{16}}\left(\frac{\varphi}{C} \right)^{\theta}\right]^{\frac{-1}{k_{10}}}-1 \right)^{-1} +\frac{k_{3}}{k_{1}} \right]^{\frac{1}{(1+\nu)}} \nonumber \\
&\times \biggl\{2(2m+1)f_{0} \nonumber \\ &-\frac{2}{\theta}(2mg_{0}+1)h_{0}k_{6}k_{10}\left( 1-\left[\frac{1}{k_{16}}\left(\frac{\varphi}{C} \right)^{\theta} \right]^{\frac{1}{k_{10}}}\right)  \nonumber \\ &  -\frac{\omega_{0}k^2_{6}k^2_{10}}{2\theta^2}
\left( 1-\left[\frac{1}{k_{16}}\left(\frac{\varphi}{C} \right)^{\theta} \right]^{\frac{1}{k_{10}}}\right)^2 \biggl\}.\label{case1-chap-59}
\end{align}
A reasonable condition compatible with (\ref{11case1-chap-57}) and (\ref{case1-chap-57}) is that we set the parameters in such a manner that  $k^{-1/k_{10}}_{16}(\varphi/C)^{\theta/k_{10}}\ll 1$. This correct assumption leads to the form of interest for the potential:
\begin{align}\label{case1-chap-60}
V(\varphi)=V_{0ch4}\varphi^{2+\frac{\theta}{k_{10}\left(1+\nu \right)}},
\end{align}
where
\begin{align}\label{case1-chap-61}
V_{0ch4}=\left(\frac{k_{3}}{C^{k_{10}}k^{k_{10}}_{16}k_{1}} \right)^{\frac{1}{\left(1+\nu \right)}} \biggl\{2(2m+1)f_{0} -\frac{2}{\theta}(2mg_{0}+1)h_{0}k_{6}k_{10}
-\frac{\omega_{0}k^2_{6}k^{2}_{10}}{2\theta^2} \biggl\}.
\end{align}
Because $\frac{\theta}{k_{10}\left(1+\nu \right)}>0$, therefore, this case of solution renders the higher orders than 2, and thus $\Gamma<1$ (`Thawing' type).\\

\noindent\pqrmybox{red}{\textbf{$\bullet$} \textbf{Comparison of the cosmological constant and the modified, generalized, and standard Chaplygin gases}}\vspace{5mm}

According to Planck data, the equation of state of dark energy is constrained to $w_{0}=-1.006 \pm 0.045$, by combining Planck data with other astrophysical data, including Type Ia supernovae. If $w$ differs from $-1$, it is likely to change with time. However, a possible way of describing this, is to consider the case of a Taylor expansion of $w$ at first order in
the scale factor, parameterized by $w=w_{0}+(1-a)w_{a}$. But, let us go another way on:\\
For the cosmological constant case, $P=-\rho$, we have $w=-1$ which is constant, and hence, if we accept that this model is true, then we must justify the little difference $-0.006 \pm 0.045$ by the Taylor expansion. But, in the `modified', `generalized', and `standard' Chaplygin gases, it is needless to use Taylor expansion because this difference is completely justifiable. The EoS parameter for the `modified Chaplygin gas' is
\begin{align}\label{case1-chap-62}
w=\sigma_{1}-\sigma_{2}\left(\frac{\sigma_{2}}{1+\sigma_{1}}
+\frac{b_{0}}{B^{(1+\sigma_{1})(1+\nu)(m+2)}} \right)^{-1}.
\end{align}
Setting the constant of integration $b_{0}$ to be very small, it may be written as
\begin{align}\label{case1-chap-63}
w=-1
+\frac{b_{0}(1+\sigma_{1})}{B^{(1+\sigma_{1})(1+\nu)(m+2)}}.
\end{align}
or equivalently
\begin{align}\label{case1-chap-64}
w=-1
\pm \frac{\left|b_{0}\right| (1+\sigma_{1})}{B^{(1+\sigma_{1})(1+\nu)(m+2)}},
\end{align}
where $b_{0}$ has been replaced by $\pm |b_{0}|$, since the sign of $b_{0}$ is undetermined. Here, such as observation data, Table (\ref{bestfitted}), we cannot say anything about $\sigma_{2}$, as it does not appear in $\omega$. Limpidly, $w$ is close to minus one because $b_{0}$ and additionally $\sigma_{1}$ were taken very small. As is observed, the little fluctuations around $w_{0}=-1$ are predicted automatically and because at the present time we have $B^{(m+2)}_{0}=1$, thus, $\pm |b_{0}|(1+\sigma_{1})=-0.006 \pm 0.045$ is acquired. Clearly, the generalized Chaplygin gas which is obtained by putting $\sigma_{1}=0$, gives $\pm |b_{0}|=-0.006 \pm 0.045$. Thanks to the constant of integration, $b_{0}$, it may be argued that there is no difference among the modified, generalized, and standard Chaplygin gases when one deals with $\omega$ and consequently the deceleration parameter $q$ because of the relation $3\omega= 2q-1$. Therefore, the type of Chaplygin gas has no effect on important events, namely `phase of the universe (Phantom/Quintessence)' and `status of the expansion of the universe (Accelerated/Decelerated)'.\\

\noindent\rmybox{}{\subsection{The case $f(\varphi)=f_{0}=constant.$}}\vspace{5mm}

For a constant coupling function, $f(\varphi)=f_{0}=const.$, in view of Noether symmetry approach which fixes
\begin{align}\label{Z13}
f(\varphi)=f_{0}, \qquad h(\varphi)=h_{0}, \qquad \omega(\varphi)=\omega_{0},
\end{align}
where, without loss of generality, we have put $k_{2}=1$ (See (\ref{sol for NS1})), the basic equation, (\ref{v15}), would be:
\begin{align}\label{Z14}
\lambda^{\prime}_{1}\left( \frac{M_{,BB}}{M}
+\frac{M_{,B}}{M}\frac{H_{2,B}}{H_{2}}\right)+\lambda^{\prime}_{2}\frac{1}{B}
\frac{M_{,B}}{M}
+\lambda^{\prime}_{3}\frac{1}{B}\frac{H_{2,B}}{H_{2}}=0,
\end{align}
where
\begin{align}\label{Z15}
&\lambda^{\prime}_{1}=\frac{2(2mg_{0}+1)}{(m+2)\omega_{0}},\quad
\lambda^{\prime}_{2}=\frac{-2 g_{0}(2m+1)}{\omega_{0}}, \nonumber \\ &
\lambda^{\prime}_{3}=\frac{2(2m+1)}{(2mg_{0}+1)}\frac{f_{0}}{h^{2}_{0} }.
\end{align}
Only the constant factors are different from that of the previous case studied ($f(\varphi)=f_{0}\varphi^2$), with $(const.)^{\prime}$ replacing $(const.)$.\\
The auxiliary equation, namely (\ref{vg12}), to (\ref{Z14}) turns out to be
\begin{align}\label{Z16}
M=\left(\frac{M_{0}h^2_{0}}{C^2_{0}} \right) \exp \left[\frac{\omega_{0}(m+2)}{2h_{0}(2mg_{0}+1)}(\varphi-\phi_{0}) \right],
\end{align}
where $\phi_{0}$ is an integration constant. It yields
\begin{align}\label{Z17}
\varphi=\varphi_{0} \; \ln[\theta_{1}M]+\phi_{0},
\end{align}
in which
\begin{align}\label{Z18}
\varphi_{0}=\frac{2h_{0}(2mg_{0}+1)}{\omega_{0}(m+2)}, \qquad
\theta_{1}=\frac{C^2_{0}}{M_{0}h^2_{0}}.
\end{align}
As is observed, the auxiliary equations are different between two cases (i.e. $f=f_{0}\varphi^2$ and $f=f_{0}$). Anyway, since the rest of the calculations can easily be performed in a similar way, so it does not seem worth trying again.\\

\noindent\hrmybox{}{\section{A General Prescription for $F(R)$ and $F(T)$ Gravities\label{gp}}}\vspace{5mm}

In this section, we would like to present a general prescription to make a \textit{case-study} of the reconstruction of $F(R)$ and $F(T)$ gravities and the relevant analysis. On combining the basics structure and rules from ref. \cite{ref1} and some enhancements from the current paper, fourteen steps should respectively be considered:
\begin{enumerate}
  \item \label{stage1} Find the forms of unknown functions, excluding the scalar potential, via a standard way, for example, the `Noether symmetry approach'. Note that if you want to find a general basic equation, such as this paper, then only the functions which do not depend explicitly upon the scalar field and its derivatives, must be identified.
  \item \label{stage2} Using Hamiltonian constraint equation, write down the form of the potential in terms of scalar field and its derivatives and the Hubble parameter.
  \item \label{stage3} Rewrite the form of the potential obtained in `Step \ref{stage2}' by considering all the functions as functions of the scale factor $a$ instead of time $t$.
  \item \label{stage4} Differentiate the form of the potential obtained in `Step \ref{stage2}' with respect to the scalar field $\varphi$ and specify the form of $V^{\prime}$.
  \item \label{stage5} Simplify the Klein-Gordon equation by inserting $V^{\prime}$ achieved in `Step \ref{stage4}'.
  \item \label{stage6} Perform a transformation, $\chi=\chi(\varphi,\varphi_{,a})$, in such a way that the equation created in `Stage \ref{stage5}' to be an equation in terms of $\chi_{,a}$, $\chi$, $a$, $H$ (the Hubble parameter), and $H_{,a}$. It is worth mentioning that this transformation is completely model-dependent. Finding a suitable transformation leading to an analytical solution is really impossible in some cases. Anyway, I suggest that it is better first try with a transformation like $\chi=Y_{1}(\varphi_{,a})/Y_{2}(\varphi)$ where $Y_{2}$ and $Y_{1}$ are the unknown functions in terms of the scalar field and its derivative with respect to the scale factor, respectively.
  \item \label{stage7} Make a connection between $\chi$ and a new function, $\mathcal{F}$, by the relation $\chi=C \mathcal{F}_{,a}/\mathcal{F}$ and rewrite the modified equation gained in `Step \ref{stage6}'. Now, a basic equation is attained. Note that the form of $H_{,a}/H$ in terms of the scale factor must be specified by the models, for example, Chaplygin gas.
  \item \label{stage8} Establish an auxiliary equation between the new function $\mathcal{F}$ and the scalar field $\varphi$ using two definitions of $\chi$.
  \item \label{stage9} Find the form of the new function $\mathcal{F}=\mathcal{F}(a)$ by solving the equation obtained in `Step \ref{stage7}'.
  \item \label{stage10} By having recourse to `Step \ref{stage8}' and `Step \ref{stage9}', specify the relation between the scalar field $\varphi$ and scale factor $a$, viz., $\tilde{\varphi}=\varphi(a)$ and $\tilde{a}=a(\varphi)$.
  \item \label{stage11} Obtain the form of the potential in terms of the scale factor, $\tilde{V}=V(a)$, using `Step \ref{stage3}' and `Step \ref{stage10}'.
  \item \label{stage12} Utilizing $\tilde{a}=a(\varphi)$, convert $\tilde{V}=V(a)$ to $V=V(\varphi)$.
  \item \label{stage13} Identify the domains of parameters and constants through the ``$\mathfrak{B}\text{-Function}$ Method'', and note that the common domains of parameters must be taken into account.
  \item \label{stage14} By the use of the results obtained in `Stage \ref{stage13}' and the function $\Gamma=V V^{\prime \prime}/ (V^{\prime})^2$ or equivalently $\Gamma=\mathfrak{B}[\varphi,0;V^{\prime}(\varphi)]/\mathfrak{B}
      [\varphi,0;V(\varphi)]$, specify that when does the potential $V(\varphi)$ have `Tracker' ($\Gamma>1$), `Scaling' ($\Gamma=1$), and `Thawing' ($\Gamma<1$) solutions.
\end{enumerate}

Of course, this prescription is suggested for the well-known actions of $F(R)$ and $F(T)$ gravities, but implicitly contains others as well.\\

\noindent\hrmybox{}{\section{Conclusion}}\vspace{5mm}

In this paper, the reconstruction procedure extended in ref. \cite{ref1} was utilized for the reconstruction of the potential of a generalized model of
\(F(T)\) gravity in the homogeneous backgrounds (FRW and LRS Bianchi I). A detailed investigation was performed for three well-known cases: \textit{i:} Barotropic Fluid; \textit{ii:} Cosmological Constant; \textit{iii:} Modified Chaplygin Gas. Our manners in dealing with the special functions emerged from these cases, were so different from other papers and led to the desired results especially the forms of interest for the potentials.

Pursuant to the astrophysical data, the true range of $m$, ($A=B^m$), was found that is very close to $1$, hence, unlike several papers, beyond the range (\ref{m-range}) is admittedly forbidden.

According to the correspondence principle and observational data, some corrections were added to the `Modified Chaplygin Gas' model. Furthermore, due to the unification theory, it was indicated that the constant parameter in the generalized Chaplygin gas model, namely $\nu$ in $P=-\sigma_{2}/\rho^{\nu}$, might not be absolutely constant, but could be slowly varying with time $t$. More precisely, it is a time-dependent variable falling so slowly from an initial constant value within $0.2 \leq \nu \leq 1.8$ to zero. However, it is better that we take the aforementioned initial domain of $\nu$ as $|\nu| \lesssim 0.05$ when we focus only on the dark sector. It was demonstrated that unlike the `Cosmological Constant' model in which the EoS parameter is exactly minus one, hence, the Taylor expansion series is demanded to elucidate its departures from minus one, in the `standard', `generalized', and `modified' Chaplygin gases a little fluctuations of the amount of the equation of state of dark energy around minus one, are completely justifiable without doing any further thing. Also, it was indicated that there is no difference among the modified, generalized, and standard Chaplygin gases when one has the EoS parameter (and consequently, deceleration parameter) as a subject matter.

The analysis of obtained results was carried by the $\mathfrak{B}\text{-function}$ method out. With this approach, we found out that all the constant parameters are tightly coupled, hence one does not allow to select the arbitrary amounts of the constants in the data analysis in order to justify some of the important events such as phase crossing, late-time-accelerated expansion, and etcetera. Physical domains of constant parameters were also achieved by other ways, namely using the known scale factors of different epochs of the universe, that confirmed with those which obtained using the $\mathfrak{B}\text{-function}$ method.

Finally, in order to study the gravitational actions via the reconstruction method, a general description pursuant to ref. \cite{ref1} and the current paper, including fourteen steps, was recommended.\\

\section*{\noindent\goldmybox{red}{\vspace{3mm} Acknowledgments \vspace{3mm}}}

This work has been supported financially by Research Institute for Astronomy $\&$ Astrophysics of Maragha (RIAAM) under research project No. 1/5440-35.\\

\noindent\hrmybox{}{\section{Supplement 1: Noether Symmetry Approach\label{mathematical supplement}}}\vspace{5mm}

The choice of the unknown functions, somewhat arbitrary, such as coupling functions $f(\varphi)$, $U(\varphi, \varphi_{,\mu}\varphi^{\mu})$, $\omega(\varphi)$ and the potential $V(\varphi)$ in (\ref{action}) has given rise to the objection of fine tuning, the very problem whose solutions have been set out through inflationary theories. Therefore, it is favorable to have a path to extract unknown functions especially the potential or at least some criteria for admissible potentials. One such approach is based on the Noether symmetry approach and it was applied by many authors (for example, see Refs. \cite{g51,no2,no3,no4,no5,no6,no7,no8,no9,no10,no11,no12,no13,no14,no15,no16,
no17,Y1no17,Y2no17,Y3no17,no18,no19,no20,no21,no22,no23,no24,no25,no26,no27,no28,no29,no30,no31,
no32,no33,no34,no35,no36,no37,no38}). Noether symmetry approach which is one of the most beautiful fruits of the calculus of variations, permits one to find out conserved quantities from the presence of variational symmetries \cite{g16}. However, some hidden conserved currents may not be obtained by the Noether symmetry approach \cite{g17, g31}. It is worth mentioning that beside the Noether symmetry approach, another approach, the Beyond Noether Symmetry approach (B.N.S. approach), has recently been suggested \cite{g51}. The B.N.S. method has two important properties: first, carrying more conserved currents than the Noether symmetry approach is feasible with it; second, if Noether symmetry approach fails to achieve the purpose, it is most probable that via B.N.S. one may find a suitable physical solutions. Nonetheless, the first option in considering cosmological models should be Noether symmetry approach.

In this supplement, we study the Noether symmetry approach for the action (\ref{action}). This paper is sufficiently long, hence we ignore to provide a short review of this method. We may refer the readers to study several papers of this subject, for example I recommend the papers of Prof. S. Capozziello on this subject.

The configuration space of the point-like Lagrangian (\ref{point like lagrangian}) is $Q=\{B,\varphi,T\}$ whose tangent space is $TQ=\{B,\varphi,T,\dot{B},\dot{\varphi},\dot{T}\}$. The existence of the Noether symmetry implies the existence of a vector field $\mathbf{X}$,
\begin{align}\label{X}
\mathbf{X}=\gamma \frac{\partial}{\partial B}+\alpha \frac{\partial}{\partial \varphi}+\eta \frac{\partial}{\partial T}
+\gamma_{,t} \frac{\partial}{\partial \dot{B}}+\alpha_{,t} \frac{\partial}{\partial \dot{\varphi}}+\eta_{,t} \frac{\partial}{\partial \dot{T}},
\end{align}
where
\begin{align*}
&\gamma=\gamma(B,\varphi,T)\longrightarrow \gamma_{,t} = \dot{B} \frac{\partial \gamma}{\partial B}+ \dot{\varphi} \frac{\partial \gamma}{\partial \varphi}+\dot{T} \frac{\partial \gamma}{\partial T},
\nonumber \\&\alpha=\alpha(B,\varphi,T)\longrightarrow \alpha_{,t} = \dot{B} \frac{\partial \alpha}{\partial B}+ \dot{\varphi} \frac{\partial \alpha}{\partial \varphi}+\dot{T} \frac{\partial \alpha}{\partial T},
\nonumber \\&\eta=\eta(B,\varphi,T)\longrightarrow \eta_{,t} = \dot{B} \frac{\partial \eta}{\partial B}+ \dot{\varphi} \frac{\partial \eta}{\partial \varphi}+\dot{T} \frac{\partial \eta}{\partial T},
\end{align*}
such that
\begin{align*}
&\mathfrak{L}_{\textbf{X}}L =0 \nonumber \\& \rightarrow \gamma \frac{\partial L}{\partial B}+\alpha \frac{\partial L}{\partial \varphi}+\eta \frac{\partial L}{\partial T}
+\gamma_{,t} \frac{\partial L}{\partial \dot{B}}+\alpha_{,t} \frac{\partial L}{\partial \dot{\varphi}}+\eta_{,t} \frac{\partial L}{\partial \dot{T}} =0.
\end{align*}

This condition yields the following system of linear partial differential equations
\begin{align}
&\left(\frac{\partial \alpha}{\partial T}\right)=0, \label{nseq1}\end{align}\begin{align}
&(2m+1)\left(\frac{\partial \alpha}{\partial B}\right)=0, \label{nseq2}\end{align}\begin{align}
&(2m+1)\left(\frac{\partial \alpha}{\partial T}\right)=0, \label{nseq3}\end{align}\begin{align}
&(2m+1)\left(\frac{\partial \gamma}{\partial T}\right)=0, \label{nseq4}\end{align}\begin{align}
&(2m+1)\left(\frac{\partial \gamma}{\partial \varphi}\right)=0,\label{nseq5}\end{align}\begin{align}
&\left(T g^{\tau} -g\right)\left(\frac{\partial \alpha}{\partial T}\right)=0, \label{nseq6}\end{align}\begin{align}
&\left(T g^{\tau} -g\right)\left(\frac{\partial \alpha}{\partial B}\right)=0, \label{nseq7}\end{align}\begin{align}
&(m+2)\gamma V-\alpha B V^{\prime}=0, \label{nseq8} \end{align}\begin{align}
&4(2m+1)f\left(\frac{\partial \gamma}{\partial \varphi}\right)-\omega B^2 \left(\frac{\partial \alpha}{\partial B}\right)=0, \label{9}\end{align}\begin{align}
&(m+2)\gamma \omega+\alpha B \omega^{\prime}+2\omega B \left(\frac{\partial \alpha}{\partial \varphi}\right)=0, \label{nseq10} \end{align}\begin{align}
&(2m+1)\left[m\gamma f+\alpha B f^{\prime}+2Bf\left(\frac{\partial \gamma}{\partial B}\right) \right]=0, \label{nseq11}\end{align}\begin{align}
&(2m+1)\bigg[m\gamma h g^{\tau}+\alpha B h^{\prime}g^{\tau}+\eta hB g^{\tau \tau} \nonumber \\ &+hBg^{\tau}\left(\frac{\partial \alpha}{\partial \varphi}\right)+2hBg^{\tau}\left(\frac{\partial \gamma}{\partial B}\right) \bigg]=0, \label{nseq12}\end{align}\begin{align}
&(m+2)\left(\gamma hTg^{\tau}- \gamma hg\right) \nonumber \\ &+\left(Tg^{\tau}-g \right) \left(\alpha Bh^{\prime}+hB\left(\frac{\partial \alpha}{\partial \varphi}\right)\right)+\eta hTBg^{\tau \tau}=0. \label{nseq13}
\end{align}
Utilizing the separation of variables, this system of linear partial differential equations may easily be solved. One may then obtain
\begin{align}\label{sol for NS}
&\eta=0, \quad \alpha=k_{1}\varphi+k_{2}, \quad \gamma=\frac{-2k_{1}B}{m+2}, \nonumber \\ & \omega(\varphi)=\omega_{0}, \quad g(T)=g_{0} \sqrt{-2(2m+1)T}, \nonumber \\&f(\varphi)=f_{0} \left(k_{1}\varphi+k_{2} \right)^2, \quad h(\varphi)=h_{0} \left(k_{1}\varphi+k_{2} \right), \nonumber \\ & V(\varphi)=\frac{V_{0}}{\left(k_{1}\varphi+k_{2} \right)^2}.
\end{align}
in which $f_{0}$, $\omega_{0}$, $V_{0}$, $h_{0}$, $g_{0}$, $k_{1}$ and $k_{2}$ are constants of integration. Therefore, two well-known sets are achieved as follow:\\

\begin{itemize}
 \item \noindent\qrmybox{red}{\textbf{Case 1:}}
 \end{itemize}
  This case is achieved by taking $k_{2}=0$:
\begin{align}\label{sol for NS2}
&\eta=0, \quad \alpha=k_{1}\varphi, \quad \gamma=\frac{-2k_{1}B}{m+2}, \nonumber \\ & \omega(\varphi)=\omega_{0}, \quad g(T)=g_{0} \sqrt{-2(2m+1)T},\nonumber \\&f(\varphi)=f_{0} \left(k_{1}\varphi\right)^2, \quad h(\varphi)=h_{0} \left(k_{1}\varphi\right), \nonumber \\ & V(\varphi)=\frac{V_{0}}{\left(k_{1}\varphi\right)^2}.
\end{align}
So, the corresponding conserved current is
\begin{align}\label{C1}
\mathbf{I}_{1}=&\frac{-k_{1}B^m}{m+2}\bigg[-2m|1+2m|g_{0}h_{0}
B\dot{B}\varphi\dot{\varphi} \nonumber \\&+8(2m+1)f_{0}\varphi^2\dot{B}^2+(m+2)B^2\varphi V^{\prime} \nonumber \\ &+(m+2)\omega_{0}B^2 \dot{\varphi}^2-2(m+2)B^2 V \bigg].
\end{align}

  \begin{itemize}
  \item \noindent\qrmybox{red}{\textbf{Case 2:}}
  \end{itemize}
  This case is earned by putting $k_{1}=0$:
  \begin{align}\label{sol for NS1}
&\eta=0, \quad \alpha=k_{2}, \quad \gamma=0, \nonumber \\ & \omega(\varphi)=\omega_{0}, \quad g(T)=g_{0} \sqrt{-2(2m+1)T},\\&f(\varphi)=f_{0} k_{2}^2, \quad h(\varphi)=h_{0}k_{2}, \nonumber \\ & V(\varphi)=\frac{V_{0}}{k_{2}^2}.
\end{align}
Therefore, its conserved current reads
\begin{align}\label{C2}
\mathbf{I}_{2}= -k_{2}B^{m+2}V^{\prime}.
\end{align}

It is seen that the Noether symmetry fixes the unknown functions. Limpidly, in both cases, $g(T)$ has the same form. Since in this work we want to reconstruct the model, not more, hence we ignore to proceed (We do not go further, as it leads to the exact solution). The desirable thing is that the unknown functions excluding the potential, which are inserted into basic equation (\ref{v15}), may be given via a standard approach (such as Noether symmetry approach).\\

\hrule \hrule \hrule \hrule \hrule \hrule

\end{document}